\documentclass[oldversion]{aa}  

\usepackage{fp, fp-exp}
\usepackage{xcolor}
\usepackage{graphicx}
\usepackage{txfonts}
\begin{document}

   \title{Multiepoch, multiwavelength study of accretion onto T~Tauri\thanks{Based on observations obtained with XMM-Newton, an ESA science mission with instruments and contributions directly funded by ESA Member States and NASA, and based on observations obtained by the Chandra X-ray observatory.}}

    \subtitle{X-ray versus optical and UV accretion tracers}

   \author{P. C. Schneider\inst{1}
          \and
          H. M. G\"unther\inst{2}
          \and
          J. Robrade\inst{1}
          \and
          J. H. M. M. Schmitt\inst{1}
          \and
          M. G\"udel\inst{3}
          }

   \institute{ Hamburger Sternwarte, Gojenbergsweg 112,
              21029 Hamburg, Germany
               \email{astro@pcschneider.eu}
     \and  
              Massachusetts Institute of Technology,
            Kavli Institute for Astrophysics \& Space Research,
            77 Massachusetts Avenue, Cambridge, MA 02139, USA
      \and 
          University of Vienna, Department of Astrophysics, T\"urkenschanzstrasse 17, A-1180 Vienna, Austria
         }

   \date{received; accepted}

% \abstract{}{}{}{}{} 
% 5 {} token are mandatory
 
  \abstract
   {
      Classical T~Tauri stars (CTTSs) accrete matter from the inner edge of their surrounding 
      circumstellar disks. The impact of the accretion material on the stellar 
      atmosphere results in a strong shock, which causes emission from the X-ray to 
      the near-infrared (NIR) domain.    Shock velocities of several 100\,km\,s$^{-1}$ imply that the immediate
   post shock plasma emits mainly in X-rays. Indeed, two X-ray diagnostics, the so-called soft excess
   and the high densities observed in He-like triplets, differentiate CTTSs from their non-accreting siblings.
   However, accretion shock properties derived from X-ray diagnostics often contradict established ultraviolet (UV)--NIR accretion tracers and a physical model simultaneously explaining both, X-ray and UV--NIR  accretion tracers, is not yet available.
   
   We present new XMM-Newton and Chandra grating observations of the CTTS T~Tauri combined with UV and optical 
   data. During all epochs, the soft excess is large and the densities derived from the O~{\sc vii} and Ne~{\sc ix}
   He-like triplets are compatible with coronal densities.  This confirms that the soft X-ray emission cannot originate in
   accretion funnels that carry the bulk of the accretion rate despite T~Tauri's large soft excess. Instead, we 
   propose a model of radially density stratified accretion columns to explain the density diagnostics
   and the soft excess.   
   In addition, accretion rate and X-ray luminosity are 
   inversely correlated in T~Tauri over several epochs. Such an anti-correlation  has been 
   observed in samples of stars. Hence the process causing it must be intrinsic to the accretion process, 
   and we speculate that the stellar 
   magnetic field configuration  on the visibile hemisphere affects both the accretion rate and the coronal emission, eventually causing the observed anti-correlation.

     }


   \maketitle
%
%________________________________________________________________

% Some values for T Tau N
\def \MTTau{2.2}
\def \RTTau{3.6}
\def \neO{0.43}
\def \neNe{2.5}
\def \MaccTTau{1.0}

%________________________________________________________________

\section{Introduction}
Classical T~Tauri stars (CTTSs) accrete material from their surrounding
protoplanetary disk through a mechanism referred to as magnetospheric
accretion \citep{Koenigl_1991}. Strong ($\sim$kG) stellar magnetic fields
interact with the inner
disk regions and channel disk material along the magnetic field lines onto the
stellar surface, where  a strong shock forms heating the material to 
temperatures of several $10^6$\,K. With this temperature, post-shock emission should be seen 
in the X-ray regime \citep{Calvet_1998, Guenther_2007, Schneider_2017}.
Indeed, density diagnostics from He-like triplets revealed plasma 
densities in CTTSs that are significantly enhanced compared to
purely coronal sources \citep{Kastner_2002, Stelzer_2004}. Also, accreting stars show an excess of cool plasma
compared to non-accreting stars, the so-called soft excess
\citep{Robrade_2007, Guedel_2007_soft}. Taken together, this suggests
that accretion does cause observable features in the X-ray regime.

However, the mass accretion rates derived from X-ray lines supposedly originating
in the accretion shock, for example, O~{\sc vii}, often fall short of the
rates derived from well-established accretion diagnostics
\citep{Curran_2011}. Local absorption of the X-rays has been hypothesized
to explain this discrepancy \citep{Argiroffi_2007}.
In particular, a deep grating observation of TW~Hya 
suggests that O~{\sc vii} does not trace the accretion spot directly, but
rather material splattered out of the accretion shock, while Ne~{\sc ix} 
emission is a more direct probe of the accretion spot properties in TW~Hya
\citep{Brickhouse_2010, Dupree_2012}.
X-ray emission from accretion shocks has been  the subject of many theoretical
works \citep[e.g.,][]{Guenther_2007, Sacco_2008,Orlando_2010,
Matsakos_2013, Bonito_2014}, which indicate that local absorption can indeed 
alter the observed accretion shock emission.

In addition to accretion shocks, jets driven by CTTSs can generate X-ray 
emission mainly at soft photon energies
\citep[e.g.,][]{Raga_2002, Guedel_DG, Schneider_2008,
Schneider_2011}. The emission from the jet often
suffers less absorption than that of the central source, which enhances
its contribution in systems with (strongly) absorbed driving sources. 
The plasma densities expected in jets are low compared to those expected in 
accretion shocks, which should be detectable in the He-like triplets.

In this paper, 
we present new observations of the prototypical T~Tauri, which is one of the most massive CTTSs. By studying its
accretion shock properties from X-ray to optical wavelengths, we 
wish to advance our understanding of stellar accretion.
This paper is organized as follows. We start with an introduction to the T~Tau system in Sect.~\ref{sect:T_Tau} and 
describe our data and reduction methods in Sect.~\ref{sect:Data}. Estimates of the mass accretion
rate are provided in Sect.~\ref{sect:Macc}. The results of the X-ray imaging  
and grating spectroscopy 
are provided in Sects.~\ref{sect:CCD} and \ref{sect:Gratings}, respectively. Section~\ref{sect:XvsUV}
presents the correlation between  X-ray and ultraviolet (UV) count rates and the results are discussed in Sect.~\ref{sect:disc}.
We close with a short summary and outlook in Sect.~\ref{sect:summary}.

\section{The target: T Tau \label{sect:T_Tau}}
T~Tau is likely to be a hierarchical triple system consisting of the three components 
N, Sa, and Sb. There has been some speculation of a fourth component in the 
T~Tau system \citep{Nisenson_1985, Ray_1997, Csepany_2015}, however, none of 
these features were conclusively shown to be stellar \citep{Kasper_2016}.
Hence, we assume that T~Tau is a triple system consisting of N and Sab 
in the following.

The angular separation of the northern component and
the center of mass of the southern system is about 0.7\,arcsec, while
the semi-major axis of the Sa/Sb system is only 0.085\,arcsec
\citep[see ][who also determine the masses of the two southern components
as $M_{Sa}=2.12\pm0.10\,M_\odot$ and $M_{Sb}=0.53\pm0.06\,M_\odot$]{Koehler_2016}. 
The northern component, T~Tau~N,  is viewed nearly pole-on \citep[$i=13^\circ$,][]{Herbst_1986} and
dominates the optical emission 
as well as the soft X-ray emission of the system, because it is less absorbed than the two 
southern components  \citep[$A_V\approx1.5\,$mag for N and $A_V\gtrsim15$ for
the southern components. See][]{White_2001, Calvet_2004, Duchene_2005, 
vanBoekel_2010,Herczeg_2014}. 
This differential absorption allows us to 
probe accretion onto an intermediate mass star without spatially separating
T~Tau~N from the two southern components Sa and Sb. In the following, we will 
therefore use T~Tau synonymously with T~Tau~N unless noted otherwise.
We use  a distance of $139\pm6$\,pc to T~Tau based on the GAIA parallax
\citep{Gaia1, Gaia2} to facilitate comparability with previous results, but 
we also note that the distance based on Very Long Baseline Array 
(VLBA) radio data has a lower error and formally disagrees from the 
GAIA value \citep[$147.6\pm0.6$\,pc, see][]{Loinard_2007}, and adopting the 
VLBA distance would increase the derived fluxes by 13\,\%.

T~Tau~N (spT: K0, age: $\sim1$\,Myr) is one of the most massive CTTSs with a
mass typically 
quoted as about $2\,M_\odot$, 
for example, as 2.4\,$M_\odot$ in \citet{Guedel_2007} or as 
$2.1\,M_\odot$ in \citet{White_2001,Podio_2014}. Also, T~Tau 
is known to possess a strong, large-scale magnetic field. \citet{Johns-Krull_2007}
reports an average magnetic field strength of 2.4\,kG and an equatorial
field strength of 0.8\,kG derived from studying the  Zeeman splitting of
magnetic field sensitive Ti~{\sc i}~lines in the 
near-infrared (IR). The average magnetic field strength is
similar to the
values reported for lower mass CTTSs.
Therefore, accretion onto 
T~Tau~N should proceed analogously to other magnetic CTTSs despite its high
stellar mass.  

\citet{Calvet_2004} study accretion in intermediate mass 
CTTSs and find 
an accretion rate of $\dot{M}_{acc}=4.4\times10^{-8}\,M_\odot$\,yr$^{-1}$
and an
accretion spot surface covering fraction  of 3.3\,\%  for T~Tau by 
using reddening-corrected UV fluxes and shock models, that is, the accretion 
properties of T~Tau match expectations based on lower mass CTTSs.
T~Tau N and Sa/Sb also drive outflows similar to other accreting CTTSs
\citep[see review by][]{Frank_2014}. 
The northern component is thought to be responsible
for the approximately east-west oriented outflow \citep[e.g.,][]{Eisloeffel_1998},
while the two southern components probably drive mainly north-south oriented
outflows \citep[see][for an overview of the 
system]{Loinard_2007}.

The presence of additional stellar components in a system 
can disturb protoplanetary
disks or even truncate the disk as in RW~Aur~A \citep{Cabrit_2006, Dai_2015}. 
Despite the proximity of the southern component (100\,au),
the disk around T~Tau~N is similar to other disks around single 
CTTSs \citep{Akeson_1998}. Therefore, accretion is likely unaffected by the presence of the 
southern binary and all accretion features are intrinsic to
T~Tau~N. The inner dust disk radius has been modeled to be at 0.1\,au for T~Tau~N 
\citep{Ratzka_2009} from   interferometry using the MID-infrared Interferometric instrument (MIDI) at the Very Large Telescope (VLT), while the
co-rotation radius is at 0.05\,au or 11\,$R_\odot$ = 3\,$R_\star$
using the estimated rotation period of 2.8\,days \citep{Bouvier_1995}.

From an X-ray point of view, T~Tau is quite special \citep{Guedel_2007}. 
On one hand, an 
XMM-Newton observation shows
a strong soft excess indicative of a surplus of cool plasma compared
to similar non-accreting stars. On the other hand,
T~Tau is the only CTTS with a strong magnetic
field and X-ray density diagnostics indicating low 
densities in the cool plasma, while lower mass CTTSs
consistently show high densities in X-ray density diagnostics. Only for single
HAe stars have X-ray grating diagnostics shown such
low densities in the O~{\sc vii} triplet. Therefore, T~Tau shares 
its low density X-ray diagnostics with accreting stars of similar mass, but 
its stellar magnetic field with lower mass stars. 

We obtained new X-ray data to check if
the low density in O~{\sc vii} is a stable property,
if Ne~{\sc ix} as a tracer of plasma with higher 
temperatures than O~{\sc vii} also indicates low densities, and to compare
X-ray emission with accretion rates. 
Addressing these items in comparison to T~Tau's siblings will 
help us to better understand 
the stellar accretion process.

\section{Observations and data processing \label{sect:Data}}
We present new X-ray observations of T~Tau obtained with XMM-Newton 
and Chandra (details in Table~\ref{tab:obs}) as well as 
H$\alpha$ profiles 
and photometry obtained near-simultaneously with the X-ray data.

\begin{table*}[th]
\centering
\caption{Analyzed X-ray observations.  \label{tab:obs}}
\begin{tabular}{ l l r c c}
\hline\hline
Observatory & Date & Exp. time (ks)  & Obs-ID & \\
XMM-Newton & 2005 Aug 15th & 83.0 & 0301500101 & \\
XMM-Newton & 2014 Aug 15th & 42.0 & 0744500201 & \\
XMM-Newton & 2014 Aug 25th & 44.8 & 0744500301 & \\
XMM-Newton & 2014 Sep 6th & 42.0 & 0744500401 & \\
Chandra (HETG/ACIS-S) & 2015 Jan 1st & 127.4 & 16672\\
\hline
\end{tabular}
% \tablefoottext{a}{Fixed.}
% \tablefoottext{a}{Approx. 0.05 and 0.15\,keV bins for 2013 and 2015, respectively.}
\end{table*}

\subsection{X-ray data}
We obtained three new XMM-Newton observations of
T~Tau and one new long exposure with Chandra using the
High-Energy Transmission Grating Spectrometer (HETG).
In total, four XMM-Newton and two Chandra observations
of T~Tau now exist (we omit the HRC-I image for the analysis 
since it lacks spectral information). This
X-ray coverage (see Table~\ref{tab:obs}) allows us to check for short- 
and long-term changes in the X-ray properties of T~Tau.
% Table~\ref{tab:obs} lists the X-ray observations analyzed in
% this paper.

We are mainly interested in the X-ray properties of T~Tau~N,
but the standard spectrum extraction regions
also include emission from T~Tau~S. Chandra
can marginally resolve the system, 
but not XMM-Newton. To ensure comparability between 
XMM-Newton and Chandra, we decided to keep the standard sizes of the 
extraction regions, which include both components. This procedure ensures that 
T~Tau~S contributes similarly
to the observed flux at all epochs although its
contribution is expected to be small at the relevant (i.e., soft) photon
energies. Hence,  we can directly compare the 
XMM-Newton with the Chandra data. 
We also adapt this procedure for the grating data, but 
modify the source position to the 0.2 -- 1.0\,keV centroid to
avoid wavelength shifts due to the southern component
(see Fig.~\ref{fig:Chandra_im}).

% In the RGS, the northern and southern components are not spatially
% seperated. Because this essentially also applies to the HETG spectrum
% whithout strongly degrading extraction efficiency, we decided to 
% use the standard extraction region for \emph{Chandra}, too. 

\subsection{XMM-Newton}
We use  Science Analsysis System (SAS) version 13.5 with the calibration files available in Nov. 2014
and largely follow the published 
SAS~procedures.    

The Charge-Coupled Devide (CCD)~spectra were extracted using standard procedures, which include the
screening for periods of enhanced particle background for the spectral analysis.
Light curves were corrected for detector artifacts using \texttt{epiclccorr}.

\subsubsection{RGS spectrum}
To extract spectra from the Reflection Grating Spectrometer (RGS) data, we use zeroth order positions matched to the 
0.2 -- 1.0\,keV centroid. The cross-dispersion size
includes 70\,\% of the Point Spread Function (PSF). The zeroth order image of RX~J0422.1+1934 is located 
in the RGS background region of T~Tau and is therefore excluded from the background
region.\footnote{\texttt{xmm.esac.esa.int/sas/current/.\\documentation/threads/rgs\_thread\_2.shtml.}}

Line fluxes were measured by integration within 0.085\,\AA{} around 
the nominal wavelengths. This width balances between extraction efficiency 
and background pickup. It includes about 75 -- 80\,\% of the line spread 
function according to the instrumental response and avoids the wide 
instrumental line wings. For measuring line fluxes in the grating spectra,
we do not subtract the background to preserve the Poisson nature of
the data. Our  local ``background'' thus  includes 
instrumental background, weak unresolved lines, and continuum emission. We estimate
this background 
from nearby wavelength regions without strong emission lines. As a cross-check, we
compared this background estimate with the sum of the instrumental background 
and the flux predicted from the Astrophysical Plasma Emission Code (APEC) models fitted to the CCD spectra. 
We find excellent agreement between both estimates. In wavelength regions
where both RGS\,1 and RGS\,2 have
significant effective area, we average the fluxes obtained  by both
instruments.

The effective area varies smoothly over the wavelength regions of interest.
In particular, we checked the wavelength region around the O~{\sc vii} triplet.
There, the effective area differs only by 1\,\% between the O~{\sc vii} f and i 
lines  and by 5\,\% between O~{\sc vii} r and i lines.

\subsubsection{Optical monitor: UV data \label{sect:OM}}
During the XMM-Newton observations, the optical monitor (OM) was used with 
the UVW1 filter, which has a central wavelength
of 2910\,\AA{} and a width of 830\,\AA{}. This filter choice allows us to
compare the near ultraviolet (NUV) properties of T~Tau with archival data. The OM was
operated in the ``Image Fast'' mode so that we could
generate UV light curves simultaneously to the X-ray data.

The OM detector suffers coincidence losses, which depend on the 
total detector count rate. T~Tau provides
just below 100\,cts\,s$^{-1}$, which is below the limit of
several hundred cts\,s$^{-1}$, where
the SAS automatically corrects for this effect. However, 
count rate jumps remain between the five sub-exposures that make one full OM 
image. To correct this effect, we smoothly join
subsequent sub-exposures while maintaining the mean count rate averaged 
over the observation. 
We convert the observed OM count rates into fluxes using the published
conversion factors.

\subsection{Chandra}
We used the HETG for our observation, which consists of two different gratings, the high and the medium energy gratings (HEG and MEG, respectively). Each grating
produces two spectral traces on the S-chips of the Advanced CCD Imaging Spectrometer (ACIS)  as well as a zeroth order 
image with the superb angular resolution of Chandra as 
the HETG does not degrade the point spread function (PSF).
We used the Chandra Interactive Analysis of Observations (CIAO) tools in version~4.6 \citep{Fruscione_2006} with CALDB version 4.6.3 to reduce 
the Chandra data.

\begin{figure}[t]
\centering
\hspace*{0.2cm}\fbox{\includegraphics[width=0.45\textwidth]{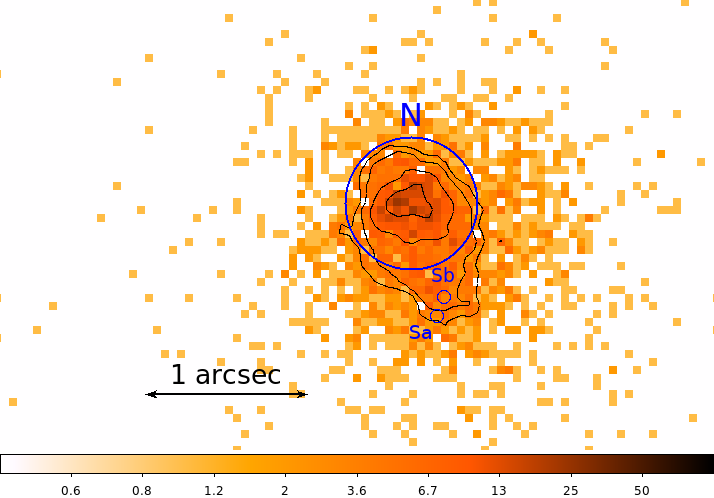}}
\hspace*{-0.18cm}\includegraphics[width=0.5\textwidth]{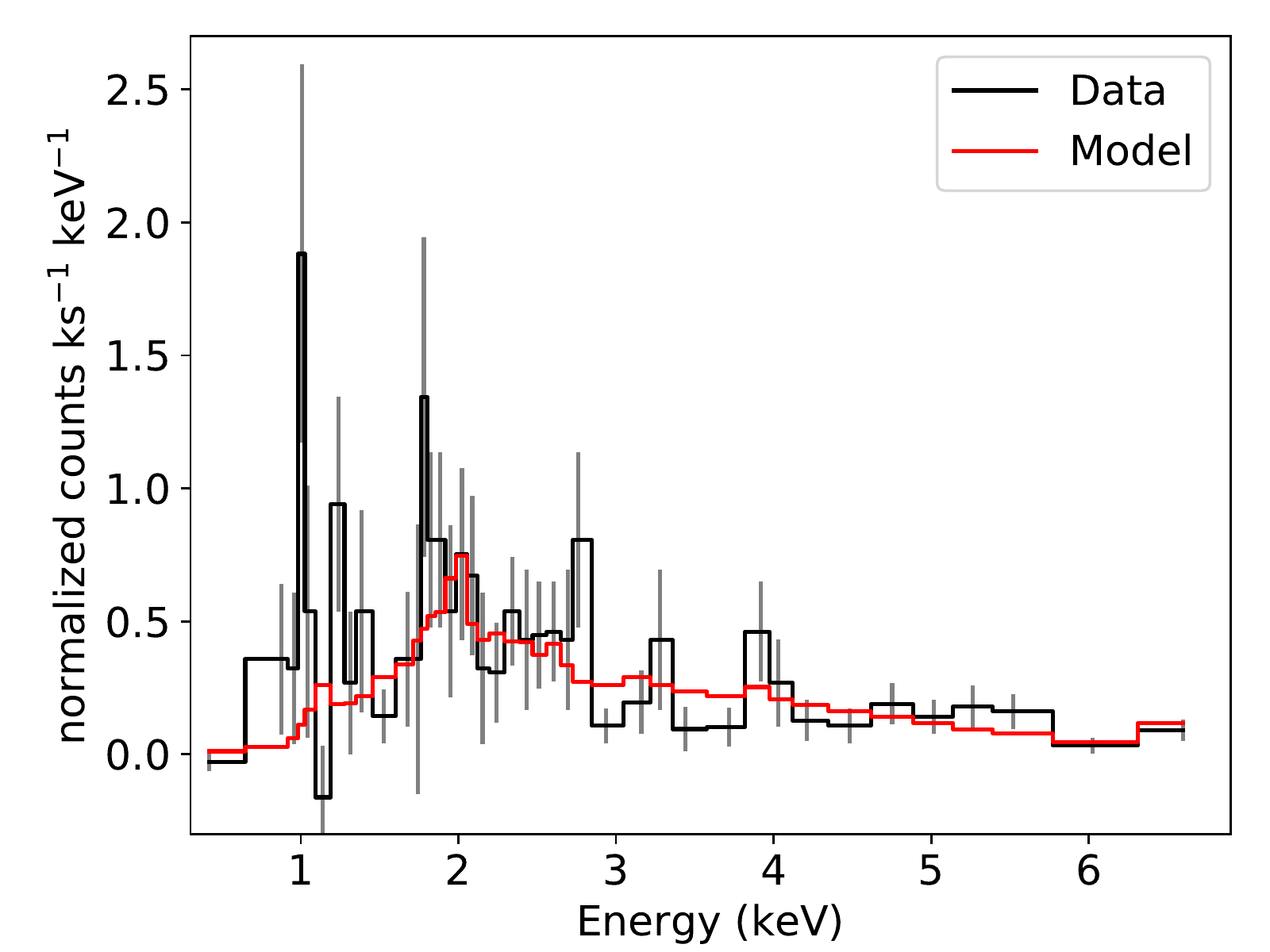}
\caption{{\bf Top}: Chandra zeroth order image of the T~Tau system
($E_{phot} = 0.3 - 5.0\,$keV).
North is up, east is left. {\bf Bottom}: 
X-ray spectrum of T~Tau~S (zeroth order HETG, i.e., ACIS-S CCD spectrum).
The red line is the 1T model.
\label{fig:Chandra_im}}
\end{figure}

\subsubsection{Zeroth order image properties and CCD spectrum of T~Tau Sab}
For our analysis of T~Tau~N, T~Tau~S is a source of contaminating photons. Therefore, 
we estimate its properties and its contribution
to the combined X-ray spectrum here. 
We extract individual spectra and light curves 
for T~Tau~N and S using small extraction regions centered on their 
nominal positions (\hbox{$r=0.8\,$pix 
($\approx0\farcs4$)} and
\hbox{$r=0.4\,$pix ($\approx0\farcs2$)} for T~Tau~N and S, respectively,
which translates to  PSF fractions of
57\,\% (N) and 26\,\% (S)). 
We experimented with different 
``background'' regions, but found only insignificant differences. 
Using the contamination by T~Tau~N estimated from a wedge-style 
region (see Fig.~\ref{fig:Chandra_im} top), we found that 
the southern 
component provides 178 net counts while 
the northern component accumulates 1690 net counts ($E_{phot} =$ 0.3 - 10.0 keV), 
which translates into a contribution fraction of 20\,\% after correction for 
the different extraction efficiencies.

Using spectra binned to five counts per bin for T~Tau~Sab (Fig.~\ref{fig:Chandra_im}, bottom) 
and \texttt{cstatistic} in xspec \citep{xspec},
we found that 
an absorbed APEC\footnote{\texttt{www.atomdb.org.}} model \citep{Foster_2012} with
$N_H=1.2_{-0.8}^{+1.1}\times10^{22}\,$cm$^{-2}$ and $kT=2.5_{-0.7}^{+2.5}\,$keV 
provides a reasonable fit to the data with a C-statistic of 53.0 for 38 degrees of freedom (corresponding 
to $\chi_{red} \approx 1.2$ with the caveat of non-Gaussian errors).
The luminosity of T~Tau Sab is $\log L_X = 30.2\pm0.24$ (0.3 -- 10.0\,keV), which is essentially the
value estimated by \citet[][$\log L_X=30.1$]{Guedel_2007}.
Such an X-ray luminosity is somewhat higher than the typical $L_X$ of young CTTSs in 
the Taurus star forming region
\citep[median $\log L_X= 29.78$, see][]{Guedel_2007_XEST}, which is likely related to
the high stellar mass T~Tau~Sa ($M_\star \gtrsim2\,M_\odot$).

Lastly, we performed a detailed spatial analysis of the zeroth order image around 
the northern component in a fashion similar to \citet{Schneider_2008} and \citet{Schneider_2011}
and conclude that 
there is no intrinsic source extension beyond a tenth of an arcsecond apart from the contamination by T~Tau Sab discussed above.

\subsubsection{HETG zeroth order CCD spectrum  of T~Tau~N+Sab}
We use a circular extraction region 
with a radius of 3\,\arcsec{} to obtain a combined (T~Tau~N + Sab) spectrum for comparison
with the XMM-Newton  European Photon Imaging Camera (EPIC) spectra. We further bin the spectrum to a minimum of 15 counts per bin 
for the subsequent spectral analysis.

\subsubsection{HETG grating spectra\label{sect:HETG}}
We reduced the Chandra data with a source position derived from
the photon centroid using the photons below 1\,keV only to remove a possible
positional shift caused by the southern component. Line counts were extracted using a 30\,m\AA-wide region around the nominal
wavelength of the line. This region includes about 98\,\% of the line spread 
function; instrumental background is negligible. Nearby, largely line-free 
regions were used to estimate the continuum and unresolved line emission,
and to derive line counts and fluxes.

\subsection{CCD fitting procedure (EPIC and HETG zeroth order)}
The combined CCD spectra (T~Tau~N plus T~Tau~Sa/Sb) are fitted with
three APEC plasma emission components with variable abundances 
(\texttt{vapec}) and photo-electric absorption (\texttt{phabs}).
In addition, we 
included a second component describing the emission from the 
southern component with the
parameters ($kT$, $N_H$, and $EM_S$) fixed to the values found in Sect.~\ref{sect:HETG}.
We also experimented with allowing a variable emission 
measure $EM_S$, but found the fit to be largely insensitive to this 
parameter. Rather, this introduced spurious effects so that we 
decided to fix all parameters of the southern component during the fitting
procedure. 

Our detailed fitting procedure is as follows: first, we assume a single value for
the absorption during all epochs and
one set of temperatures for the three emission components during all 
epochs, while allowing the $EM$ for each epoch and temperature component
to vary freely. 
Since we do not have sufficiently well-exposed grating
spectra for all epochs, we use three groups of abundances (elements with a low First Ionization Potential (FIP):
Fe, Mg, Si, Al, Ca, Ni; mid-FIP: C, N, O, S; high-FIP: Ne, Ar).
Assuming that the temperatures of the three emission components 
are equal in all five epochs is an unrealistic simplification, but 
allows us to easily investigate the 
evolution of the cool, intermediate, and hot components. 
In a second step, we release this 
assumption and allow the three temperatures to vary between the 
five epochs while still requiring the abundances and absorbing 
column density to be equal in all epochs (releasing the 
constraint on $N_H$ does not provide additional conclusive results, 
especially in regards to changes in absorbing column density between 
epochs).
This approach for fitting the CCD spectra is slightly different 
from the method employed by \citet{Guedel_2007}; in their fit to 
the RGS data,  individual 
elements were allowed to vary freely. Nevertheless, the results are largely
compatible.

\subsection{Line fitting procedure}
We obtained line fluxes by direct integration within wavelength
ranges adjusted to the respective spectral resolution of the 
instrument (see above). To measure line ratios
in the He-like triplets, we used
a model with the physical quantities as fit parameters, that is,
total flux and line ratios (e.\,g., G- and R-ratios) instead of individual line 
counts. Thereby, we avoid the pitfalls of propagating non-Gaussian errors 
and can directly obtain meaningful errors for the line ratios.
This model also restricts the line ratios explored during the 
fitting procedure to physical ranges. We use the C-statistic \citep{Cash_1979}
for the fitting.
In consequence, the line ratios (see Table~\ref{tab:grating_ratios}) do not 
necessarily reproduce exactly the ratios 
resulting from dividing the counts obtained by integration of the spectra 
(Table~\ref{tab:line_counts}); minor changes can be caused by
the broad line wings included in the fitting procedure but 
neglected when integrating counts. 

\subsection{Ground-based data}
To support the X-ray observations, we also obtained optical photometry
and H$\alpha$ spectroscopy.

\begin{figure}[t!]
\centering
\includegraphics[width=0.48\textwidth]{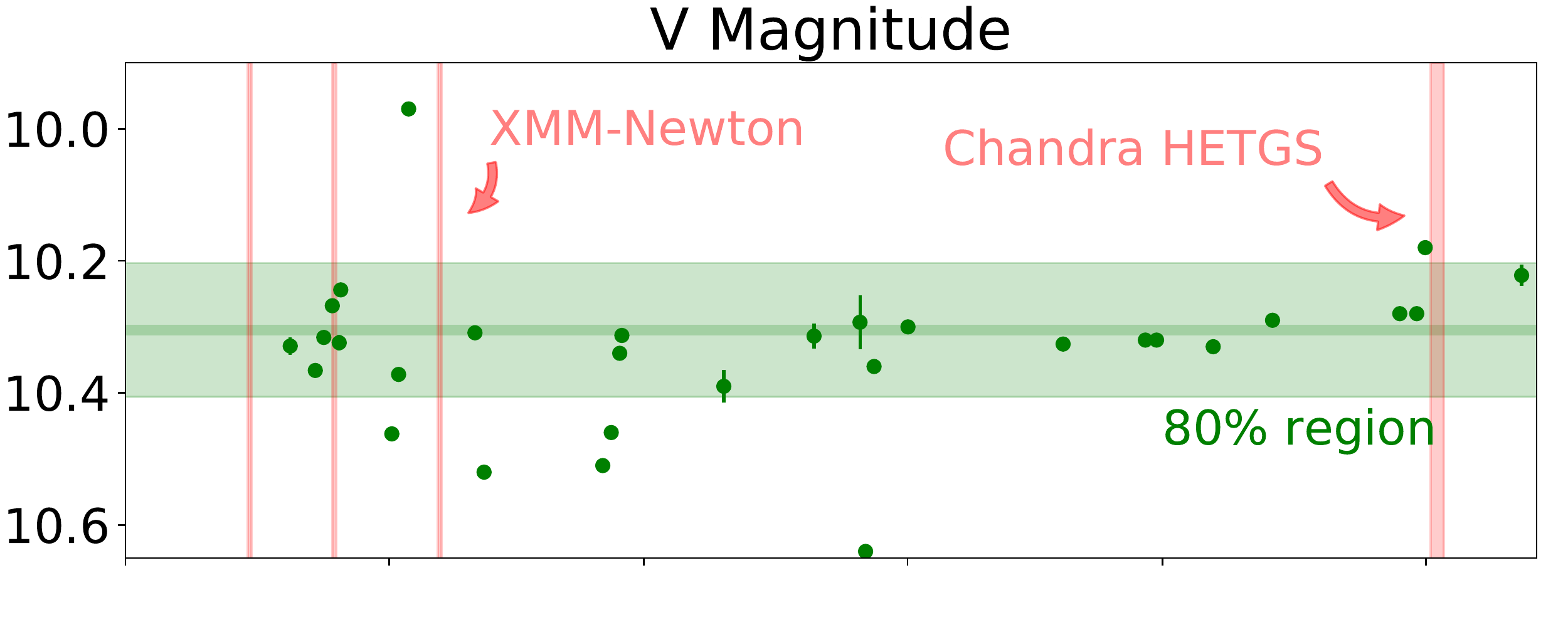}
\includegraphics[width=0.48\textwidth]{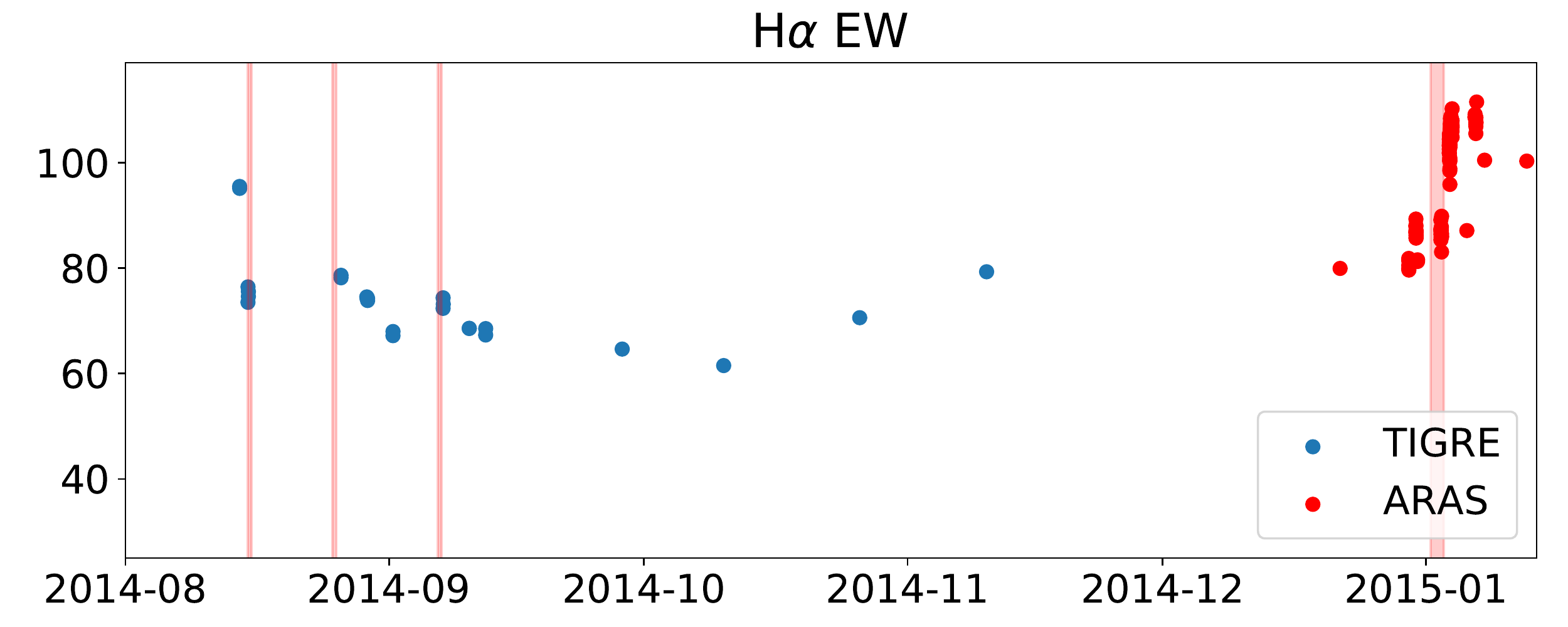}
\caption{{\bf Top}: V magnitude of the T~Tau system (unresolved) from AAVSO. The 
magnitude range containing 80\,\% of the data is marked. The vertical lines indicate
the dates of the X-ray observations.
{\bf Bottom}: H$\alpha$ EW.\label{fig:timeline}}
\end{figure}

To investigate the optical behavior of T~Tau around the X-ray observations, 
we retrieved V band magnitudes from the 
American Association of Variable Star Observers (AAVSO, Henden, A.A., 2013, 
Observations from the AAVSO International Database, http://www.aavso.org)
and provide them in Fig.~\ref{fig:timeline}~(top).
The optical brightness of T~Tau is relatively constant and we estimate
V=$10.3\pm0.1$ during the X-ray observations. This is about 30\,\% dimmer
than the mean observed by the Research Of Tracers Of Rotation (ROTOR) project \citep[V$\approx9.9$ with
a standard deviation of 0.1\,mag, see ][]{Grankin_2007}.

Spectroscopic data were 
obtained  by the Telescopio Internacional de Guanajuato Rob\'otico Espectrosc\'pico
\citep[TIGRE\footnote{\texttt{www.hs.uni-hamburg.de/DE/Ins/HRT/index.html.}},][]{Schmitt_2014}
and through the Astronomical Ring for Access to Spectroscopy (ARAS).\footnote{\texttt{www.astrosurf.com/aras/intro/intro.htm.}}
TIGRE is a 
fiber-fed spectrograph in La Luz, Mexico. The fiber entrance diameter is 
3$\arcsec$ and the spectral resolution is
\hbox{$R\approx20.000$};
data reduction procedures are described in \citet{Mittag_2010}. Due to 
instrumental issues of the TIGRE instrument around the Chandra
observation, we can only use the \hbox{H$\alpha$ 10\,\%~width} but not the Equivilent Width (EW) 
for this epoch. 

% Thus, we use additional continuum
% normalized H$\alpha$ spectra provided by ARAS  for the time around
% the \emph{Chandra} observation. 

\section{Mass accretion rates\label{sect:Macc}}
We now obtain estimates for the mass accretion rate based on established
diagnostics so that we can later compare the X-ray diagnostics with these 
accretion tracers.

\subsection{H$\alpha$ line}
\begin{figure}[t!]
\centering
\includegraphics[width=0.49\textwidth]{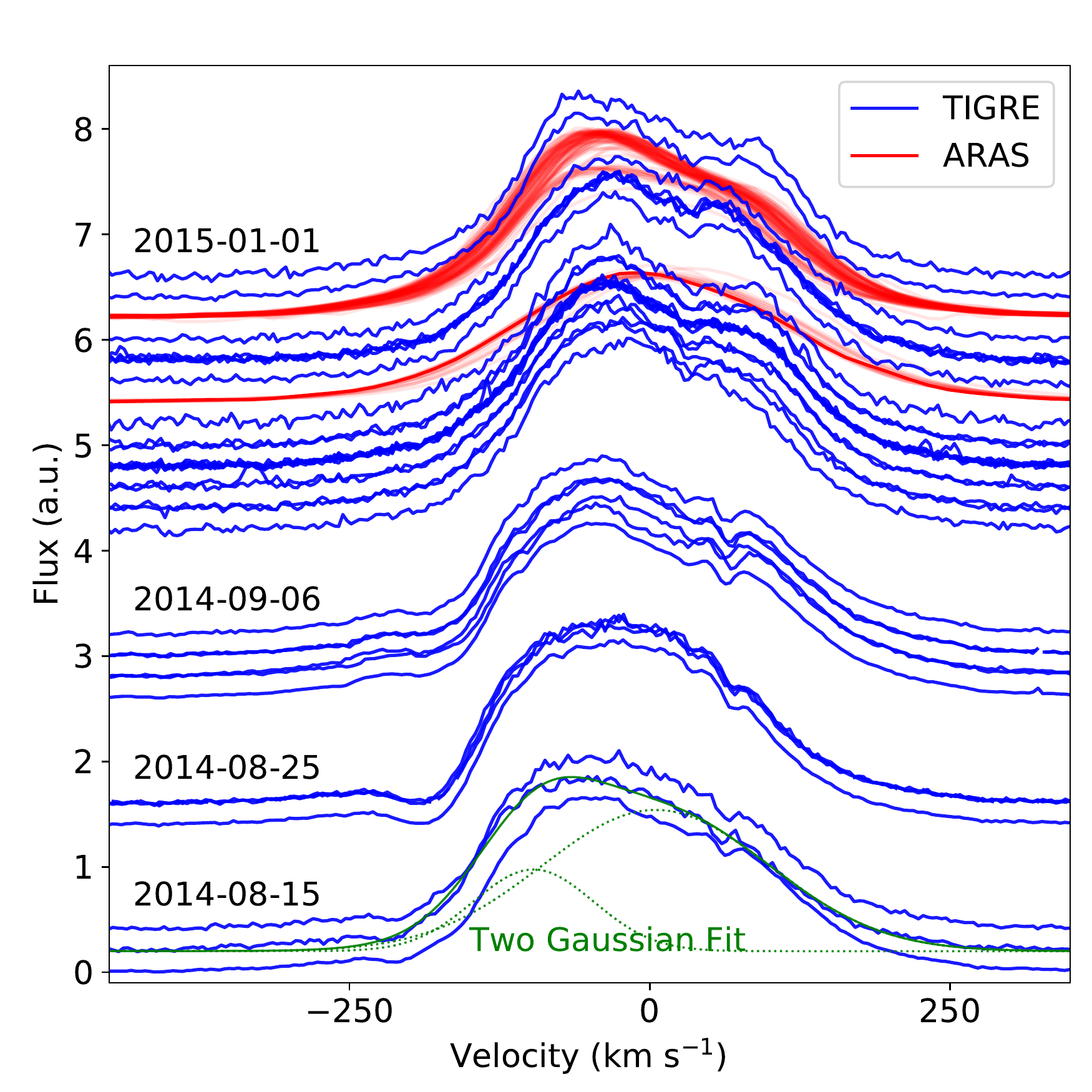}
\caption{H$\alpha$ profiles (near-) simultaneous to the X-ray observations. 
Data from ARAS has a light copper tint and is distinguishable 
from the TIGRE data by its lower spectral resolution. \label{fig:Ha_atlas}}
\end{figure}

We start our analysis with the H$\alpha$ emission, one of the most widely 
used accretion tracers. The observed profiles 
are shown in Fig.~\ref{fig:Ha_atlas}.
The overall shape of
the H$\alpha$ profiles remained relatively unchanged  between 
August 2014 and January 2015. 
Differences in equivalent width (EW) are 
$\lesssim10\,$\%, including the EW calculated from the ARAS 
data for the time of the Chandra observation.
We provide the average values calculated for the times around the X-ray observations in Table~\ref{tab:Ha}, which 
compare well with the  values observed 
by \citet{Costigan_2014} during their 2001 observations
($EW_{H\alpha}^{2001}\approx 71\,$\AA). 

The H$\alpha$ profile appears asymmetric with more emission at negative
velocities. However,  decomposition into two Gaussians shows that
only 15\,\% of the 
H$\alpha$ flux originates in a rather narrow, blue-shifted component 
($v_{peak}\approx-70\dots-90$\,km\,s$^{-1}$,
$\sigma\approx$40\,km\,s$^{-1}$), while the majority of the emission
can be explained by a broad (100\,km\,s$^{-1}$), slightly red-shifted 
($v\approx10$\,km\,s$^{-1}$) Gaussian component. The narrow, blue-shifted component
is likely associated with the known jet \citep[HH~155, e.g.,][]{Eisloeffel_1998}, while the majority
of the H$\alpha$ emission is related to the accretion process. Therefore, 
it is reasonable to use the H$\alpha$ luminosity to estimate the accretion 
luminosity despite the predominately blue-shifted velocities.

The H$\alpha$ line properties have been shown to
correlate with excess emission caused by accretion \citep[UV excess or veiling; see references in][]{Mendigutia_2015}. These relations typically originate from CTTS 
observations and have been extended to higher mass stars by simultaneously
measuring accretion and line luminosities
\citep[e.g.,][]{Mendigutia_2011}. We provide estimates for the mass accretion below 
using three different diagnostics (H$\alpha$ EW, H$\alpha$~10\,\% width, NUV flux).

\subsubsection{Accretion rates from H$\alpha$ equivalent width}

The conversion 
from H$\alpha$ luminosity to accretion rate is considered one of the more reliable
accretion diagnostics \citep[at least for CTTSs; see discussion in][]{Alcala_2014}. We convert the measured EW to 
H$\alpha$ luminosity using the measured V magnitude 
(V$\approx10.3$; see Fig.~\ref{fig:timeline} top), the
average  V-R from the \hbox{ROTOR} program \citep[][V-R $\approx$ 1.1]{Grankin_2007}
and $A_V\approx1.8$ \citep{Calvet_2004, Herczeg_2014}. This results in 
a continuum surface flux near H$\alpha$ of $6\times10^6\,$erg\,cm$^{-2}$\,s$^{-1}$\,\AA{}$^{-1}$, which
% ,
% which equals approximately the value of 
% $5\times10^6\,$erg\,cm$^{-2}$\,s$^{-1}$ assumed by \citet{Costigan_2014}
% based on spectral type and stellar radius. It also 
corresponds well
with the continuum flux published by \citet[][ we read 
$\approx1.5\times10^{-12}\,$erg\,s$^{-1}$\,cm$^{-2}$\,\AA$^{-1}$ from
their Fig.~9, while our assumed value corresponds to
$2\times10^{-12}\,$erg\,s$^{-1}$\,cm$^{-2}$\,\AA$^{-1}$]{Calvet_2004}.
We are therefore confident that the error on our estimated surface flux is less than 10\,\%. 
We then convert the derived H$\alpha$ luminosity ($\log L_{H\alpha} / L_\odot \approx -1$)
to accretion luminosity using the \citet{Mendigutia_2012} relation 
and find  
% 94 * 4e-13 / (10^(-0.4 * 0.75 *1.5)) * 4 * pi *(140 * 3.1e18)^2 = 2.51e32
% log( 94 * 4e-13 / (10^(-0.4 * 0.75 *1.5)) * 4 * pi *(140 * 3.1e18)^2 / 3.85e33) = -1.19
% Correcting for Outflow: 
% log( 94*0.8 * 4e-13 / (10^(-0.4 * 0.75 *1.5)) * 4 * pi *(140 * 3.1e18)^2 / 3.85e33) = -1.28
%
% log Lacc = 2.28 + 1.09 * log L_Ha
%2.28 + 1.09 * -1.19 = 0.9829
% Correcting for outflow: 2.28 + 1.09 * -1.28 = 0.88
%
% or 3.14 + 1.48 * -0.42 = 1.82
% or 3.14 + 1.48 * (-0.42 - 0.1) = 1.71
an accretion rate of $7.8\times10^{-7}\,M_\odot\,$yr$^{-1}$. 
% M_-8 = Lacc_30 / 820 * R / M
% 10^0.98 * 3.8e33 / 1e30 / 820 * 3.5 / 2.0 * 1e-8 = 8e-7
% log (10^0.98 * 3.8e33 / 1e30 / 820 * 3.5 / 2.0 * 1e-8) = -6.11
%
Correcting for the outflow emission ($\approx15$\,\%)
to the measured H$\alpha$ EW 
and for chromospheric H$\alpha$ (on the order of EW=1\,\AA) lowers 
the accretion rate to about $6.6\times10^{-7}\,M_\odot\,$yr$^{-1}$.

\subsubsection{Accretion rates from H$\alpha$ 10\,\% width}
Another measure for the accretion rate is the H$\alpha$ 10\,\% width
and we find an average 10\,\% width of 543\,km\,s$^{-1}$. Variability 
in the 10\,\% width is only on the few percent level for observations falling within
the XMM-Newton observing window. TIGRE data obtained with the 
replacement camera during the Chandra observation
show, on average, a similar 10\,\% width to the other data, but
with a slightly larger scatter of 20\,\%.

\begin{figure}[t!]
\includegraphics[width=0.49\textwidth]{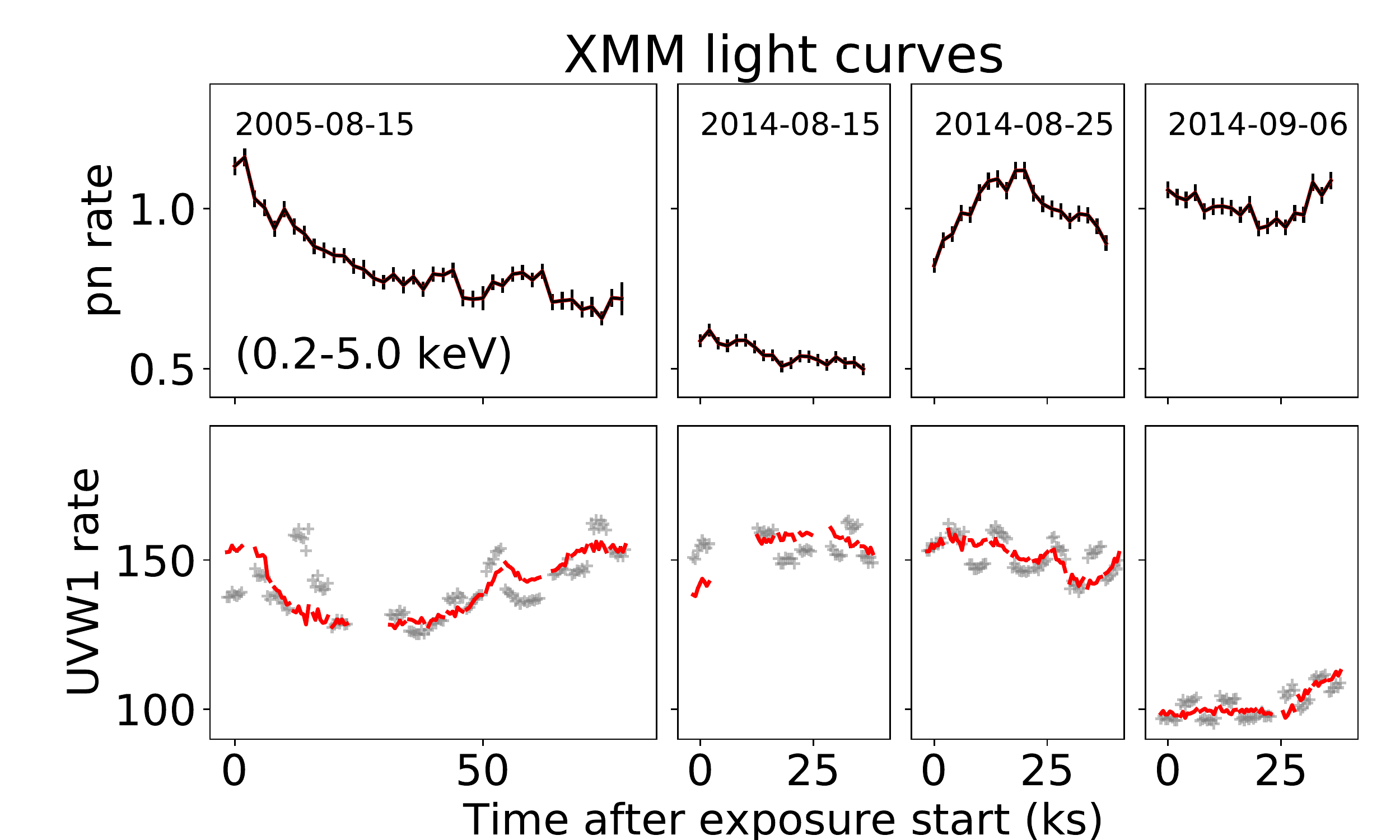}
\includegraphics[width=0.49\textwidth]{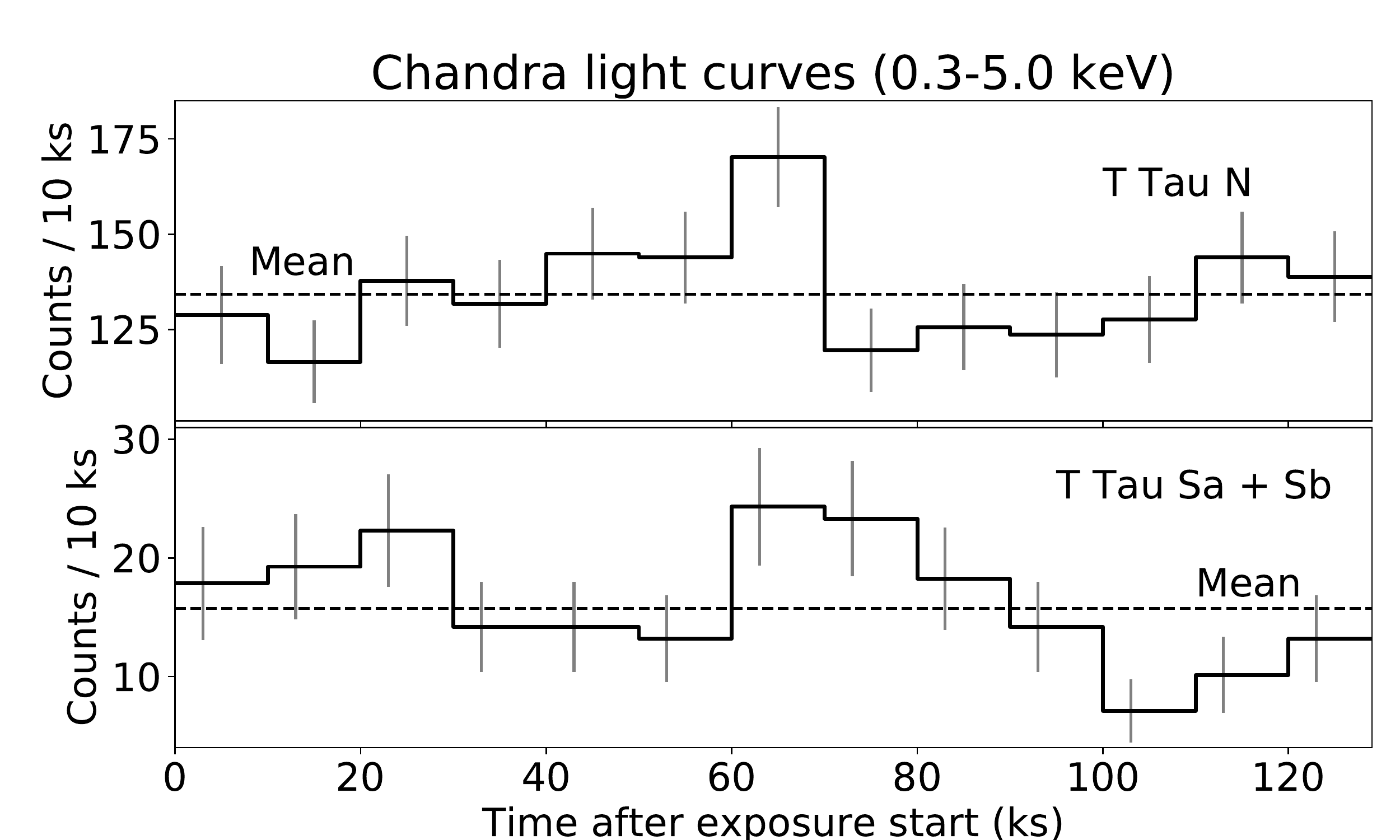}
\caption{{\bf Top}: XMM-Newton X-ray and UV light curves. For the OM, the gray data points indicate the pipeline values before correcting for variable coincidence losses (for details, see Sect.~\ref{sect:OM}).
         {\bf Bottom}: Chandra X-ray light curves. \label{fig:lcs}}
\end{figure}

To convert the 10\,\%~width to accretion rate we use the relation presented by \citet[][]{Calvet_2004}, which  is based on 
comparing the accretion generated continuum with simultaneous H$\alpha$ data for a set of intermediate-mass stars. We find that the accretion rates are mostly in the range  $4-8\times10^{-8}\,M_\odot\,$yr$^{-1}$
(only the small width observed on 2014~Sep~6th  corresponds to accretion 
 rates $<10^{-8}\,M_\odot\,$yr$^{-1}$). The average width (543\,km\,s$^{-1}$) corresponds
to $2.4\times10^{-8}\,M_\odot$\,yr$^{-1}$.
% , which is an order of magnitude below the value derived from the H$\alpha$ EW.

%------------------------------------------------------------
% Ha luminosities:
%https://www.gemini.edu/cgi-bin/sciops/instruments/michelle/magnitudes.pl?magnitude=10.8&wavelength=6.5&filter=Johnson+R&option=magnitude
% %R=9.2 -> 3.632e-12 W/m^2/micron --> 3.6e-13 erg/s/cm^2/AA
%R=7.3 -> 2e-11 W/m^2/micron --> 2e-12 erg/s/cm^2/AA
% or
%http://www.astronomy.ohio-state.edu/~martini/usefuldata.html
% 217.7 * 10^(-0.4*9.2) -> 4.5e-13
% 217.7 * 10^(-0.4*7.3) -> 2.6e-12 # 
%------------------------------------------------------------

% % http://www.astronomy.ohio-state.edu/~martini/usefuldata.html

\subsection{Accretion rates from NUV fluxes}
We list in Table~\ref{tab:UV_Fluxes} the mean fluxes
observed by the OM during the individual epochs. For $A_V=1.8$,
we find emitted (dereddened) fluxes between $1.0$ and 
$1.7\times10^{-12}$\,erg\,s$^{-1}$,
which are close to the value reported by 
\citet[][$F_{NUV}\approx10^{-12}\,$erg\,s$^{-1}$]{Calvet_2004}. 
These authors show that most of the flux at NUV wavelengths
is due to accretion. Assuming constant extinction, this indicates 
that changes in accretion luminosity
are less than a factor of two during single exposures as well as between the four 
epochs. Assuming a linear relation between NUV luminosity and  mass accretion 
rate, the UV fluxes observed by the OM indicate 
$\dot{M}_{acc}\approx4.4 \dots 7.5 \times10^{-8}\,M_\odot\,$yr$^{-1}$
\citep[][derive $\dot{M}_{acc}\approx4.4\times10^{-8}\,M_\odot\,$yr$^{-1}$ for 
       $F_{NUV}\approx10^{-12}\,$erg\,s$^{-1}$]{Calvet_2004}.

\subsection{Discussion of the different accretion rates \label{sect:accretion_rate}}

The H$\alpha$ EW gives about an order of magnitude higher accretion rates than 
the estimates
from H$\alpha$ 10\,\% width and NUV fluxes.
Since our value from the EW is similar to previous estimates also based on EW
\citep[][$\dot{M}_{acc}=1.4\times10^{-6}\,M_\odot\,$yr$^{-1}$]{Costigan_2014},
we consider the discrepancy as inherent to the employed method. The H$\alpha$ EW, while  
qualitatively correlating well with accretion, also shows a large scatter with respect
to other tracers. Since 10\,\% width as well as NUV excess, which directly traces 
the accretion shock emission, suggest similar rates, we assume in the following  
$\dot{M}_{acc}=5\times10^{-8}\,M_\odot\,$yr$^{-1}$ as representative. 
This value is rather typical for (high-mass) CTTSs \citep{Calvet_2004}. In addition,
the observed trend between stellar mass and accretion rate with $\dot{M}_{acc} \propto M^{2.1}$ 
implies $\dot{M}_{acc}=7\times10^{-8}\,M_\odot\,$yr$^{-1}$ for T~Tau \citep[see][]{Hartmann_2016},
which is very close to the derived values given the observed scatter. 
In summary, T~Tau's mass accretion rate and its minor variability are
rather typical for stars of its mass and age \citep{Venuti_2014}.

Using Eq.~1 from  \citet[][]{Guedel_2007}, we use our estimate for the mass 
accretion rate to estimate the post-shock density of an homogeneous accretion spot covering 10\,\% of the stellar 
surface (i.e., surface covering fraction 0.1) and find
$n_e=2\times10^{12}$\,cm$^{-3}$, which is in the sensitivity range of the
X-ray density diagnostics
with the O~{\sc vii} and Ne~{\sc ix} He-like triplets.

\section{General plasma properties from X-ray CCD spectra \label{sect:CCD}}
We start our investigation of the X-ray properties of T~Tau with 
the light curves before moving to the plasma properties derived from 
CCD spectroscopy.

\begin{table*}[t]
% \centering
           \caption{CCD spectral fits with 90\,\% confidence ranges. Abundances are with respect to \citet{Anders_1989}. 
                     Abundances provided for each FIP group for 
                    the 2005~Aug~15th  RGS data 
                    are numerical averages of the elements  fitted by \citet{Guedel_2007}. The discrepancy between 
                    the $N_H$ derived from the 2005~Aug~15th CCD ($3.4\times10^{21}$\,cm$^{-2}$) and RGS data 
                    ($4.9\times10^{21}$\,cm$^{-2}$) was already noted by \citet{Guedel_2007}.
           \label{tab:CCD_specs}}
            \setlength{\tabcolsep}{0.1cm}
 \begin{tabular}{ l l l l l l l l l}
  \hline\hline
     Paramater & Unit & Value & \\
     \hline
                                        & & \citet{Guedel_2007}   &             \multicolumn{5}{c}{Date (ObsID)}  \\
                                        & &     (from RGS data)       &             \multicolumn{1}{c}{2005 Aug 15th} & \multicolumn{1}{c}{2014 Aug 15th} &      \multicolumn{1}{c}{2014 Aug 25th} & \multicolumn{1}{c}{2014 Sep 6th}& \multicolumn{1}{c}{2015 Jan 1st}\\
                                        & &                      &             \multicolumn{1}{c}{(0301500101)}       &  \multicolumn{1}{c}{(0744500201)} & \multicolumn{1}{c}{(0744500401)} & \multicolumn{1}{c}{(0744500401)} & \multicolumn{1}{c}{(16672)}\\
$N_H$ & $10^{22}\,$cm$^{-2}$                    & $0.49_{-0.06}^{+0.1}$   & \multicolumn{5}{c}{$ 0.33_{-0.02}^{+ 0.02} $} \\
$T_1$ & $10^6\,$K                               &$1.76_{-0.82}^{+0.4}$   & \multicolumn{5}{c}{$ 1.91_{-0.09}^{+ 0.10} $} \\
$T_2$ & $10^6\,$K                               &$7.42_{-0.51}^{+0.45}$  & \multicolumn{5}{c}{$ 8.33_{-0.27}^{+ 0.26} $} \\
$T_3$ & $10^6\,$K                               &$28.8_{-2.3}^{+3.5}$    & \multicolumn{5}{c}{$35.6_{-1.5}^{+ 1.6} $} \\
$EM_1$ & $10^{53}\,$cm$^{-3}$                   &$9.3_{-5.5}^{+16.5}$    &  $ 0.42_{-0.42}^{+ 0.60} $&$ 5.13_{-1.11}^{+ 1.37} $&$ 5.94_{-1.40}^{+1.74} $&$ 4.52_{-1.17}^{+ 1.49} $&$ 6.81_{-3.08}^{+ 3.75} $ \\
$EM_2$ & $10^{53}\,$cm$^{-3}$                   & $4.4_{-1.6}^{+3.3}$     &$ 5.63_{-0.63}^{+ 0.66} $&$ 4.08_{-0.38}^{+ 0.39} $&$ 5.76_{-0.62}^{+ 0.64} $&$ 7.39_{-0.75}^{+ 0.77} $&$ 5.23_{-0.69}^{+ 0.21} $ \\
$EM_3$ & $10^{53}\,$cm$^{-3}$                   & $5.0_{-0.9}^{+0.6}$     &$ 4.31_{-0.21}^{+ 0.21} $&$ 1.38_{-0.10}^{+ 0.10} $&$ 4.44_{-0.20}^{+ 0.20}$&$ 3.68_{-0.21}^{+ 0.21} $&$ 2.66_{-0.21}^{+ 0.21} $ \\
% $N_{H,\,S}^a$ & $10^{22}\,$cm$^{-2}$                    & -- & \multicolumn{5}{c}{1.2}\\
% $T_S$\tablefootmark{a} & $10^6\,$K              & -- & \multicolumn{5}{c}{23.9}\\
% $EM_S$\tablefootmark{a} & $10^{53}\,$cm$^{-3}$  & -- & \multicolumn{5}{c}{$0.02$} \\
Low FIP &                                       & 0.32   & \multicolumn{5}{c}{$ 0.10_{-0.01}^{+ 0.01} $} \\[-0.1cm]
     \tiny{(Fe, Mg, Si, Al, Ca, Ni)}\\
Mid FIP  &                                       & 0.35 & \multicolumn{5}{c}{$ 0.37_{-0.04}^{+ 0.04} $} \\[-0.1cm]
\tiny{(C, N, O, S)}\\
High FIP &                                      & 0.82 & \multicolumn{5}{c}{$ 0.90_{-0.07}^{+ 0.08} $} \\[-0.1cm]
\tiny{(Ne, Ar)}\\
$\log L_X$ &erg\,s$^{-1}$ & 30.9 &      $31.00\pm0.02$ & $30.87\pm0.03$ & $31.10_{-0.02}^{+0.03}$ & $31.09_{-0.02}^{+0.03}$ & $31.02\pm0.06$ \\
    \tiny{(0.3 - 10.0 keV)}\\
         $\chi^2$ / dof      &                  &    416.4 /  382                 &  \multicolumn{5}{c}{4321.7 / 3878 }\\    
mean pn rate & cts/s & & $0.691\pm0.004$  & $0.434\pm0.004$ & $0.792\pm0.005$ & $0.803\pm005$\\             
        \hline
  \end{tabular}            
%   \\
%   \tablefoottext{a}{Fixed for the fit.}
\end{table*}

\subsection{Light curves and spectra}

The X-ray light curves (Fig.~\ref{fig:lcs})
reveal almost a factor of two difference between the epochs, but no 
obvious flares, and we use a single spectral model 
for each epoch; only the 2005 epoch shows some variability, but
we decided to fit a single model to preserve comparability with 
the results presented by \citet{Guedel_2007}.

\begin{figure*}[t]
\centering
\includegraphics[width=0.75\textwidth]{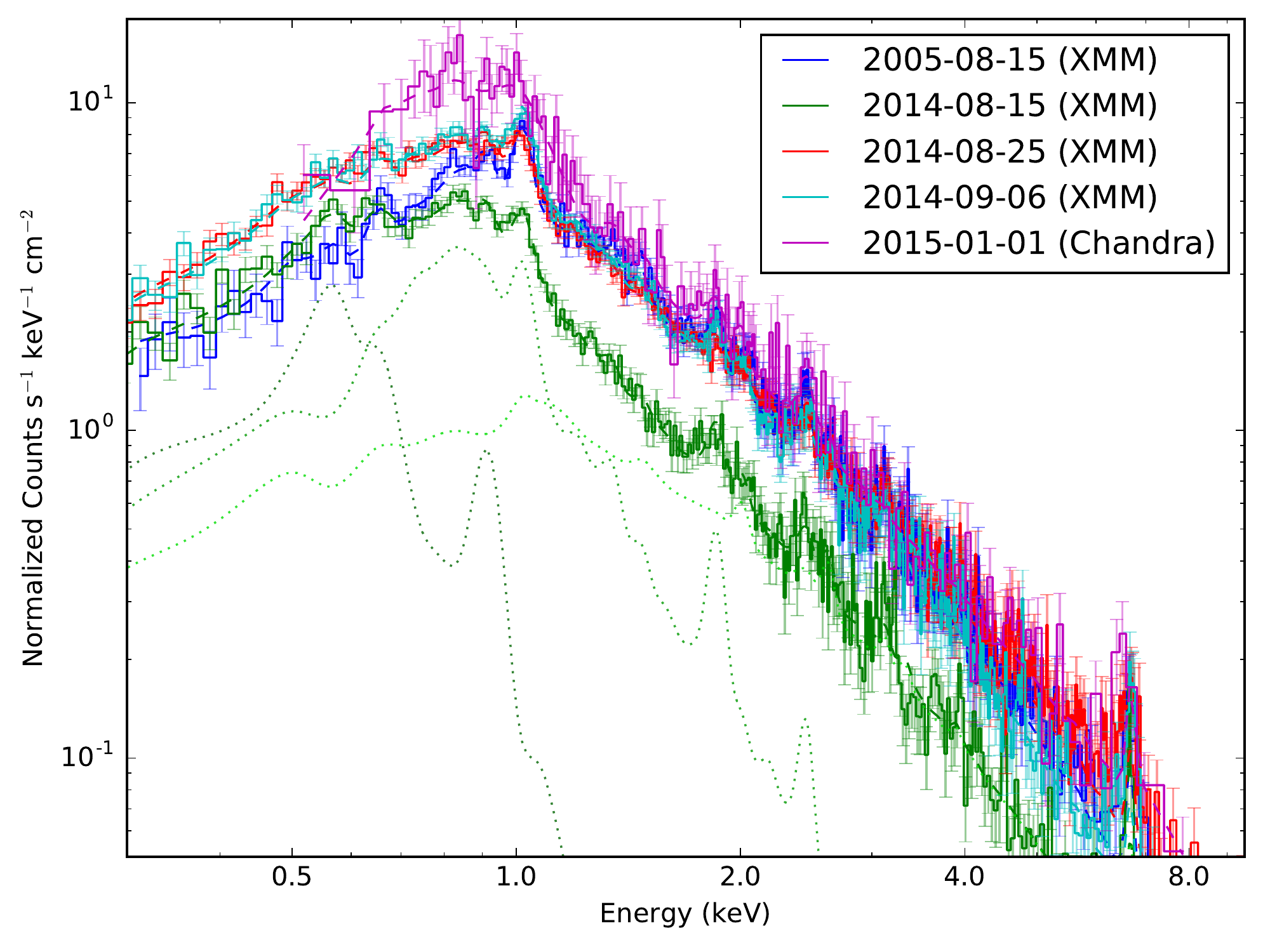}
\caption{CCD X-ray spectra of T~Tau.
The extraction region encompasses T~Tau~N and S to enable
a direct comparison of the spectra. XMM-Newton EPIC spectra have 
been merged for this plot. For illustration purposes, the dotted green lines indicate the contributions
of the individual plasma components for the 2014~Aug~15th epoch.
\label{fig:CCD_specs}}
\end{figure*}

Figure~\ref{fig:CCD_specs} shows  the CCD spectra.
Differences are more pronounced in the soft band 
($E_{phot}\leq1.0\,$keV) than in the hard spectral range 
($E_{phot}>1\,$keV), where four out of five epochs show 
similar fluxes. The lowest fluxes are observed for the 2014~Aug~15th
epoch by a factor of two. 
Ten days later, the fluxes are back to the ``normal'' level.

We experimented with several fit approaches, for example,
a single temperature grid for all epochs or variable temperatures.
We found that  the fit results obtained with a single temperature grid 
for all epochs already provides  a decent description
of the data ($\chi_{red} \approx 1.11$) while more complex models only
mildly improve the fit ($\chi_{red} \approx 1.08$). Therefore, we provide
these values in Table~\ref{tab:CCD_specs} to ease inter-comparison
between the epochs.
Differences between the various fit approaches are mainly in the
EM of the individual components, while temperatures and abundances
remain  relatively stable, for example, temperature changes are mainly in 
the 10-30\,\% range. Specifically, the coolest component's   temperature is
 stable
(we find $T_{cool} = 1.7\times10^6\,$K with typical errors of 0.1--0.2$\times10^6$\,K, i.e., slightly 
 lower than the $1.9\times10^6\,$K obtained from the joint fit).
Notable is the low $EM_1$ for 2005~Aug~15th  in the fit using
a single temperature grid that is not mirrored by a particularly 
weak O~{\sc vii} emission in the corresponding RGS spectrum.
Closer inspection shows that the particularly low $EM_1$ in 
Table~\ref{tab:CCD_specs} for 2005~Aug~15th  is partly based on 
a true lack of soft emission (about a factor of two lower than in the brighter epochs, cf. Fig.~\ref{fig:CCD_specs}), but mainly due to the cross-talk between the 
lowest and the intermediate temperature components: the intermediate temperature component is 
responsible for most of the flux at low energies and there is little 
space for emission from the low temperature component except for isolated lines like
O~{\sc vii}.
Allowing the fit to vary the temperatures freely between 
 the epochs changes 
$EM_1$/$EM_2$ from 1:13 to 1:2, which is more in line with the RGS spectrum where strong O~{\sc vii} 
emission is seen. 
The low $EM_3$ for 2014~Aug~15th, however, is clearly visible in the CCD spectra where
the lowest fluxes are observed above $E_{Phot}\approx0.6\,$keV and remains essentially
unchanged in the free temperature fit.
Inspection of the 
Fe~complex around $E_{Phot}\sim6.7\,$keV further suggests a higher Fe abundance
of the hottest plasma component during all epochs, and weak Fe~K$\alpha$ 
emission in one epoch. 

The expected post-shock plasma temperature is below 5\,MK for the free fall velocities 
$<600$\,km\,s$^{-1}$, the maximum amenable for T~Tau~N (see Table~\ref{tab:comparison}). Therefore, 
the two hotter temperature components must be associated with
coronal emission and are  
responsible for 70 to 80\,\% of the total X-ray flux (0.3 -- 10.0\,keV, 
depending on epoch). 
They indicate magnetic activity as expected 
given the measured $\sim$kG magnetic field \citep{Johns-Krull_2007}.
Thus, the X-ray spectrum of T~Tau above $0.5\,$keV
appears similar to low-mass
CTTSs and differs from young but magnetically 
inactive A~type stars of similar stellar mass like $\beta$~Pic, which 
generate orders of magnitude weaker X-ray emission 
\citep{Guenther_2012}.

The cool plasma component $T_1$ has a typical luminosity of about 
$2.6\times10^{30}$\,erg\,s$^{-1}$ and dominates the observed emission 
only at photon energies around O~{\sc vii} ($E_{Phot}\approx0.6$\,keV) and below
during some epochs (see Fig.~\ref{fig:CCD_specs}). 
It dominates the O~{\sc vii} emission by at least 
one order of magnitude while the Ne~{\sc ix} emission ($E_{Phot}\approx0.9$\,keV) originates in both
the cool and the intermediate temperature components (ratio about 1:2).

In summary, the CCD spectra indicate plasma properties similar to 
those found in active, accreting stars  \citep[cf. the XEST results,][]{Guedel_2007_XEST}.
The observed X-ray fluxes ($6-10\times10^{30}\,$erg\,s$^{-1}$ for 
the 0.3 - 10.0\,keV band) correspond
to $\log L_X / L_\star = -3.5$ 
\citep[for $L_\star = 7.5\,L_\odot$, see][]{White_2001},
typical for very active stars \citep[][]{Brickhouse_2010, 
Robrade_2007, Guenther_2007, Telleschi_2007} and 
published relations  between 
stellar mass and X-ray luminosity \citep{Telleschi_2007} suggest
$\log L_X \approx30.88$, which is within a factor of two of the measured values.

Also, the abundance pattern known as the inverse first ionization potential effect (iFIP) 
and the variability (factor two, no flares) are similar to previous X-ray
observations of accreting intermediate-mass stars.
In summary, the X-ray properties from CCD spectroscopy of T~Tau 
are similar to those observed for other (massive) CTTSs.

\begin{figure*}[t!]
\centering
\includegraphics[width=0.99\textwidth]{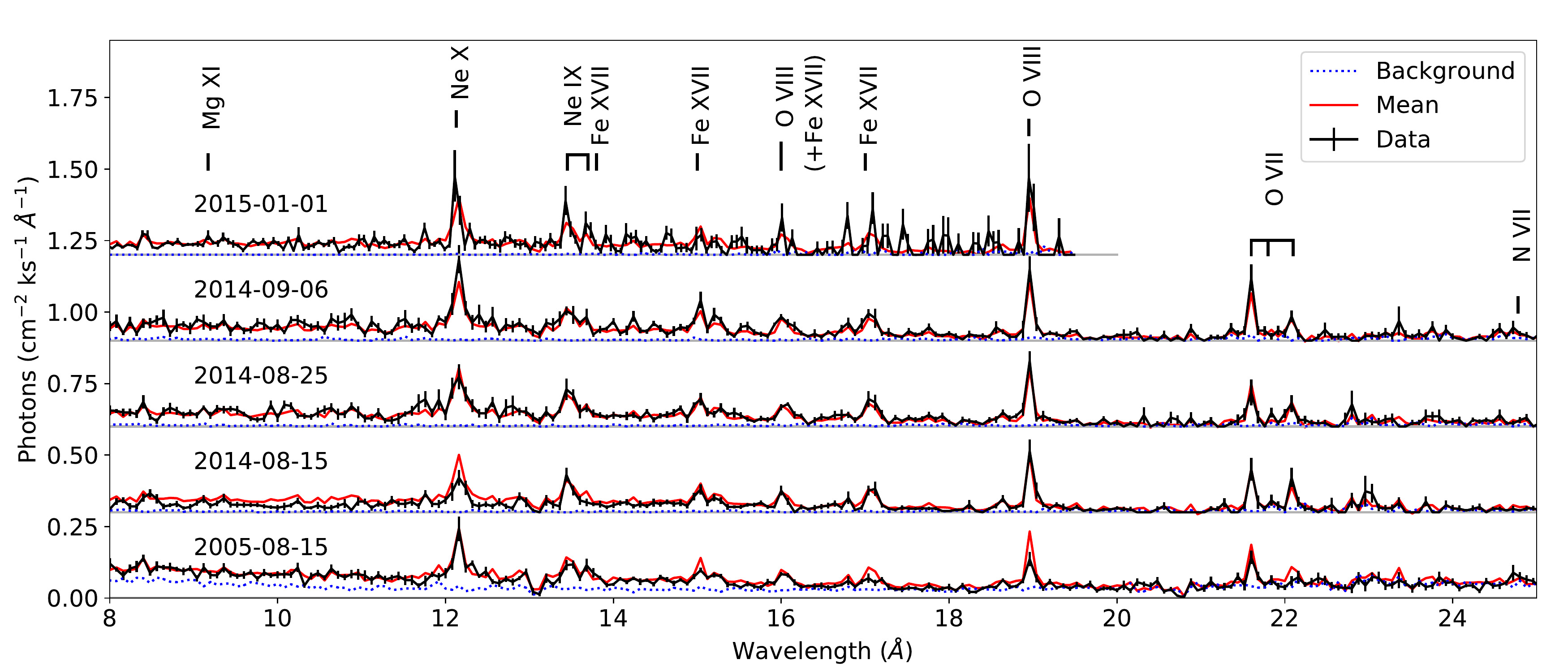}
\caption{Atlas of X-ray grating spectra. ``Mean'' indicates the mean observed spectrum. \label{fig:gratings}}
\end{figure*}

\section{Line ratios from grating spectra \label{sect:Gratings}}
Figure~\ref{fig:gratings} shows the RGS and HETG grating spectra.
We concentrate on a number of interesting diagnostic lines, but do not perform
a detailed spectral modeling. Our observing strategy with multiple 
shorter exposures results in a comparably low number of reliably measurable lines so 
that we prefer to obtain plasma properties and absorbing column density
from the CCD spectra. 
Important line fluxes are provided in
Table~\ref{tab:line_counts}. Because these lines are at quite low
energies  ($E_{phot}<0.9$\,keV), contamination by T~Tau~S is 
negligible.

In the following, we focus on He-like triplets. The
terms r, i, and f indicate the resonance, intercombination, and forbidden 
lines. Their line ratios depend on temperature (G-ratio: (f+i)/r) and 
on a combination of density and local UV field (R-ratio: f/i). 
Higher densities as well as high UV fluxes decrease the R-ratio, 
that is, larger ratios correspond to lower densities and low UV fluxes. 
In CTTSs, the photospheric UV field is considered too weak to significantly affect 
the R-ratio, and high density measurements have been regularly interpreted as
high plasma densities \citep[e.g.,][]{Telleschi_2007}.

\begin{figure}[t!]
\centering
\includegraphics[width=0.48\textwidth]{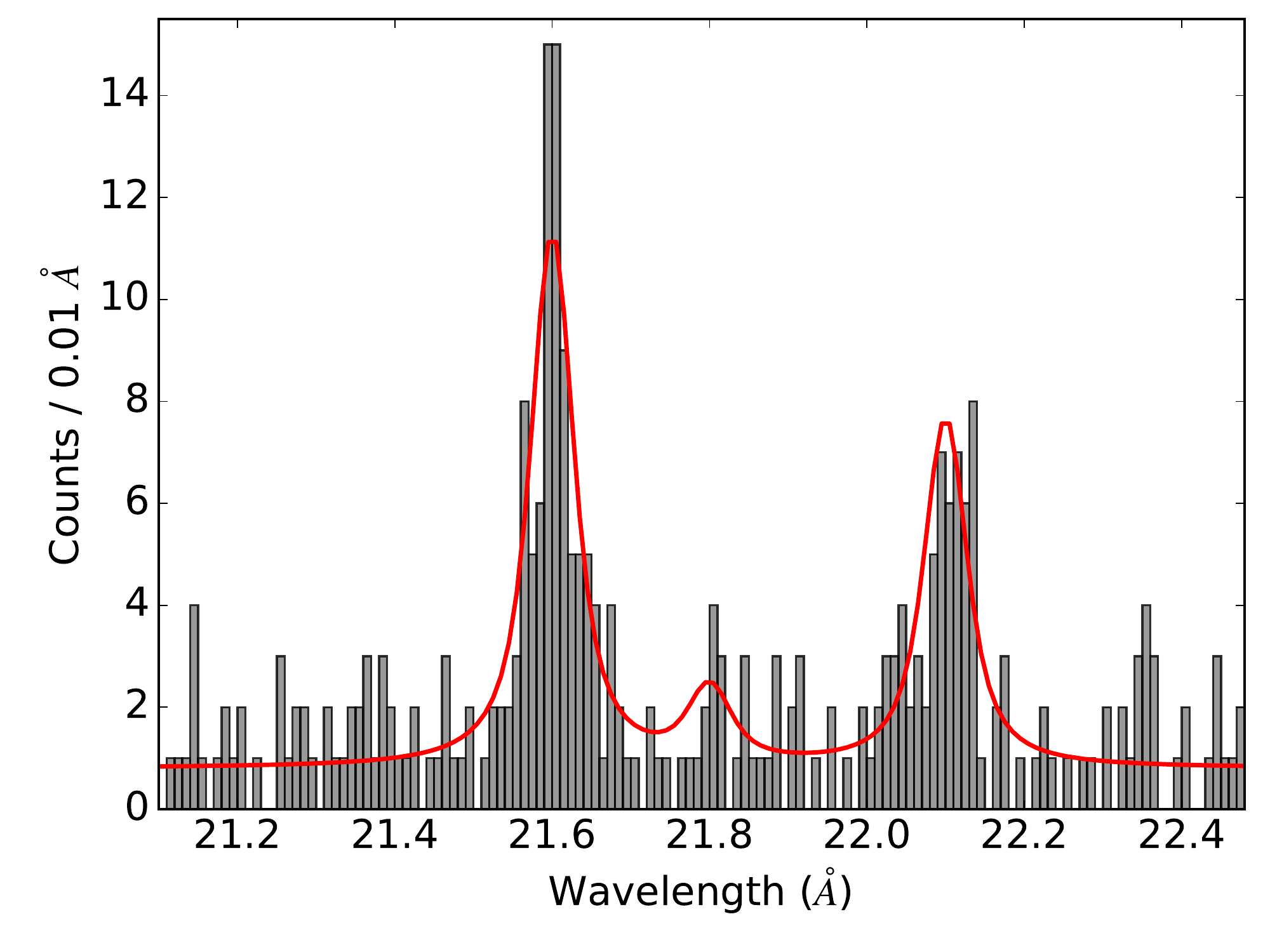}
\caption{Merged RGS spectra around the O~{\sc vii} triplet.
\label{fig:O7}}
\end{figure}

\subsection{Oxygen \label{sect:G_ratio}}
The oxygen lines provide us with important information 
on the lowest temperature plasma possibly emitted by the immediate post-shock plasma. They are 
measured from the XMM-Newton RGS spectra since too few photons were recorded in
this region in the Chandra HETG observation.

In principle, the G-ratio provides us with information on the temperature of 
the O~{\sc vii} emitting plasma. The 
measured values (between 0.6 and 1.2) 
indicate temperatures in the range between 
$\log\,T=6.0\dots6.7$, however, the 
90\,\% credibility ranges extend from 
$\log T < 6.0$ to $\log T > 7.0$. Therefore, we
use $T_1$ from the CCD 
spectroscopy, which is close to the 
peak formation temperature ($\log T = 6.3$). The 
predicted line fluxes are in good agreement with the measured values 
\citep[using the CHIANTI database, see][]{Young_2009}. Hence,
we consider $T_1$ as a good estimate of the O~{\sc vii} emitting 
plasma fully consistent with the measured G-ratios.

\subsubsection{R-ratio: Density of the O~{\sc vii}-emitting plasma \label{sect:O7-R}}
Figure~\ref{fig:gratings} 
shows that the forbidden line is stronger than the intercombination 
line during all epochs. The detailed results from line fitting listed in 
Table~\ref{tab:grating_ratios} reveal that the lower limit on the f/i-ratio 
is about two while the formal upper limit corresponds to the theoretical
limit. The 90\,\% confidence limits are $\log n_e<10.5 \dots 10.9$ for the individual epochs,
which is similar to the upper limit found  by \citet{Guedel_2007} for the 2005 
observation. This demonstrates that  the 
densities during the 2005 XMM-Newton observation are rather typical for T~Tau.

Coadding the four epochs of RGS
data results in $\log\,n_e < 10.0$  and $<10.4$ for the 1\,$\sigma$ and
90\,\% confidence ranges, respectively.
The corresponding densities depend slightly on the plasma temperature and
we have assumed a temperature of $\log T=6.3$ for the above estimates
(about $T_1$ from our CCD fits and compatible with the G-ratio).
Varying the plasma temperature by $\pm0.2$\,dex 
changes these limits by less than
0.1\,dex while simultaneously reducing the O~{\sc vii} 
emissivity by a factor of two. The upper limit is lower than the density 
estimated in Sect.~\ref{sect:accretion_rate}, and it is even lower than the 
post-shock density of an accretion funnel carrying the
derived accretion rate and covering
the entire stellar surface (such an artificial shock would have $n_e=2\times10^{11}\,$cm$^{-3}$). Therefore,
other processes must be invoked to explain the O~{\sc vii} 
emission.

\begin{table*}[t]
\centering
\small
\caption{Observed line fluxes in units of $10^{-6}$\,counts\,s$^{-1}$\,cm$^{-2}$. Numbers in brackets 
give the number of detected photons (sum of RGS\,1/RGS\,2 and positive/negative grating orders for RGS and MEG, respectively) with 
background estimate. We note that Ne~{\sc ix} fluxes from RGS spectra include contamination from Fe {\sc xvii}/{\sc xix}
and should be regarded as upper limits.\label{tab:line_counts}}
            \setlength{\tabcolsep}{0.1cm}
\begin{tabular}{ l c rrrrrrrrrrrr}
\hline\hline
Line & Wavelength & \multicolumn{2}{c}{2005~Aug~15th} & \multicolumn{2}{c}{2014~Aug~15th} & \multicolumn{2}{c}{2014~Aug~25th} &  \multicolumn{2}{c}{2014~Sep~9th} & \multicolumn{2}{c}{2015~Jan~1st}\\
 & (\AA) & \multicolumn{2}{c}{RGS} & \multicolumn{2}{c}{RGS} & \multicolumn{2}{c}{RGS} & \multicolumn{2}{c}{RGS} & \multicolumn{2}{c}{MEG} \\
\hline
Ne {\sc ix} r & 13.45  & $  7.3 \pm   2.3$ & ($ 33 -  13.5$) & $ 11.9 \pm   3.5$ & ($ 19 -   3.7$) & $  9.1 \pm   3.2$ & ($ 18 -   5.5$) & $  7.6 \pm   3.6$ & ($ 19 -   9.2$) & $ 12.3 \pm   3.9$ & ($ 24 -   2.1$)\\ 
Ne {\sc ix} i & 13.55  & $  4.8 \pm   2.1$ & ($ 27 -  14.2$) & $  4.2 \pm   2.5$ & ($  9 -   3.6$) & $  5.6 \pm   2.8$ & ($ 13 -   5.3$) & $ -0.6 \pm   2.5$ & ($  8 -   8.7$) & $  5.6 \pm   2.8$ & ($ 12 -   1.9$)\\ 
Ne {\sc ix} f & 13.70  & $  9.9 \pm   2.5$ & ($ 42 -  15.2$) & $  0.5 \pm   1.7$ & ($  4 -   3.4$) & $  3.5 \pm   2.5$ & ($ 10 -   5.1$) & $  2.3 \pm   2.8$ & ($ 11 -   8.1$) & $  8.3 \pm   3.5$ & ($ 15 -   1.6$)\\ 
O~{\sc viii} & 16.01  & $  5.2 \pm   1.6$ & ($ 55 -  19.1$) & $  4.4 \pm   2.1$ & ($ 21 -   6.6$) & $  3.6 \pm   2.0$ & ($ 22 -   9.5$) & $  4.6 \pm   2.2$ & ($ 24 -   8.8$) & $  9.5 \pm   5.6$ & ($  7 -   0.4$)\\ 
O {\sc vii}  & 18.63  & $  2.3 \pm   1.7$ & ($ 24 -  14.0$) & $  4.7 \pm   2.5$ & ($ 13 -   3.1$) & $  4.9 \pm   2.7$ & ($ 16 -   5.1$) & $  3.9 \pm   2.6$ & ($ 13 -   4.9$)& \multicolumn{2}{c}{--}\\ 
O {\sc viii} & 18.97  & $ 13.4 \pm   3.0$ & ($ 66 -  12.3$) & $ 22.4 \pm   4.9$ & ($ 49 -   2.9$) & $ 23.0 \pm   4.9$ & ($ 53 -   3.8$) & $ 26.0 \pm   5.3$ & ($ 57 -   4.2$) & $ 19.8 \pm  13.4$ & ($  7 -   0.2$)\\ 
O {\sc vii} r & 21.60  &$  10.8 \pm   2.9$ & ($ 25 -   5.7$) &$  14.8 \pm   4.4$ & ($ 13 -   0.7$) &$  11.7 \pm   3.8$ & ($ 11 -   0.7$) &$  21.5 \pm   5.3$ & ($ 19 -   1.3$)& \multicolumn{2}{c}{--}\\ 
O {\sc vii} i & 21.80  &$   1.0 \pm   1.8$ & ($  8 -   6.2$) &$   2.8 \pm   2.3$ & ($  3 -   0.8$) &$   2.6 \pm   2.1$ & ($  3 -   0.8$) &$   0.8 \pm   2.0$ & ($  2 -   1.4$)& \multicolumn{2}{c}{--}\\ 
O {\sc vii} f & 22.10  &$   8.0 \pm   2.7$ & ($ 19 -   5.4$) &$  10.6 \pm   3.9$ & ($  9 -   0.7$) &$   8.7 \pm   3.4$ & ($  8 -   0.7$) &$   8.8 \pm   3.7$ & ($  8 -   1.1$)& \multicolumn{2}{c}{--}\\ 
N {\sc vii} & 24.78  & $  2.3 \pm   1.4$ & ($ 10 -   4.4$) & $  0.0 \pm   1.1$ & ($  1 -   1.0$) & $  0.1 \pm   1.0$ & ($  1 -   0.9$) & $  1.7 \pm   1.7$ & ($  3 -   1.2$)& \multicolumn{2}{c}{--}\\ 
\hline
\end{tabular}
% \tablefoottext{a}{Fixed.}
% \tablefoottext{a}{Footnote example.}
\end{table*}

\begin{figure}[t!]
\centering
\includegraphics[width=0.48\textwidth]{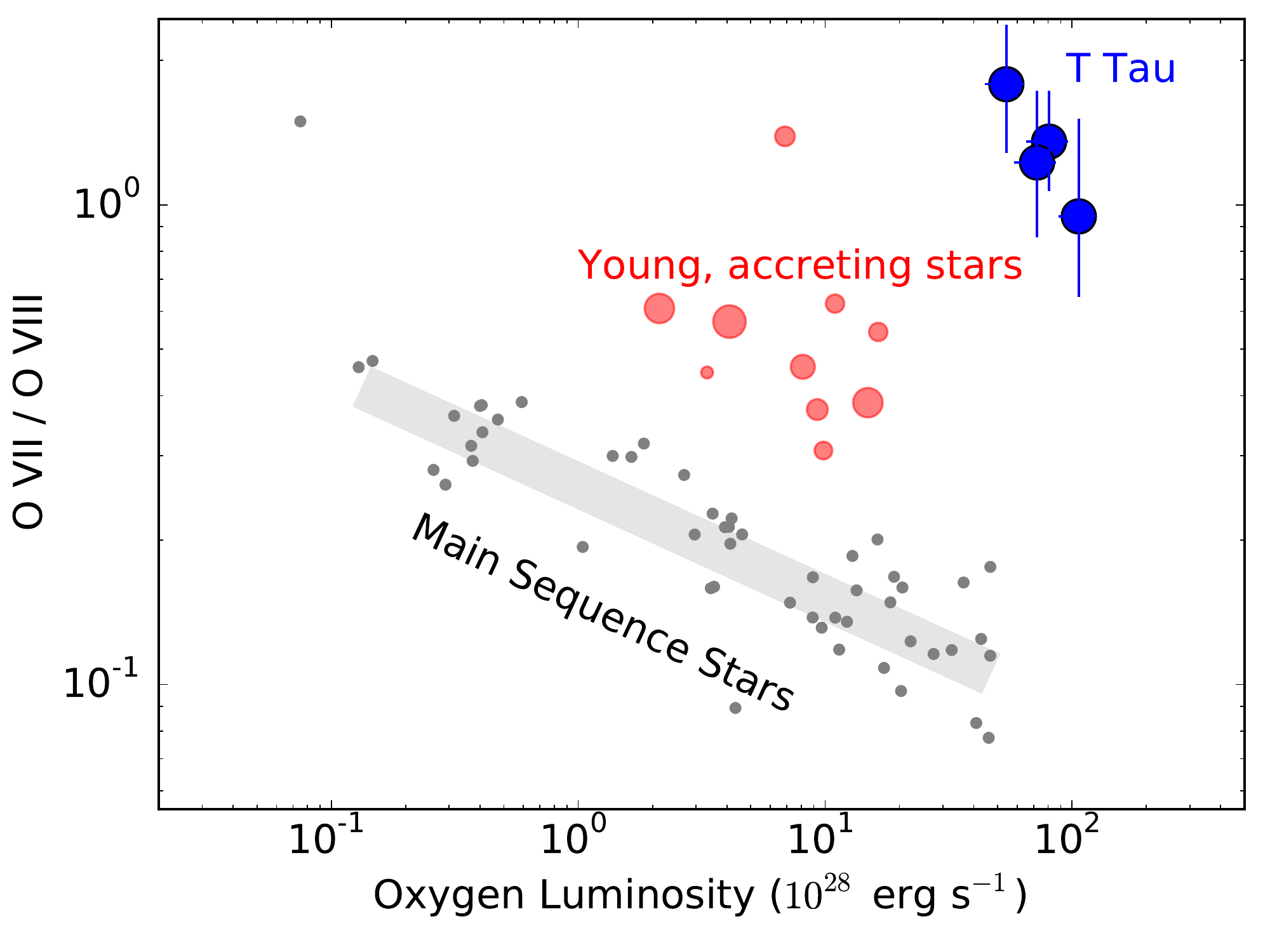}
\caption{Soft excess of T~Tau compared with other T~Tauri stars as
well as main sequence stars. Main sequence stars are mainly from 
\citet{Ness_2002, Ness_2004}.  The decrease of the O~{\sc vii} 
to O~{\sc viii} ratio with increasing oxygen luminosity is a result
of the increasing average coronal temperature for increasingly 
active MS stars \citep[e.g.,][]{Johnstone_2015}. Plot symbol sizes scale with 
mass accretion rate. 
\label{fig:soft_excess}}
\end{figure}

\subsubsection{Soft excess}
The soft excess is defined as the dereddened ratio between 
the O~{\sc vii} r and O~{\sc viii} Ly\,$\alpha$ line fluxes 
\citep{Guedel_2007_soft,Robrade_2007}. 
It depends on the assumed absorbing column 
density. In Appendix~\ref{sect:abso_O7}, we show 
that, in contrast to
TW~Hya, we do not find a different absorption 
towards the O~{\sc vii} emitting region and towards the 
corona. Therefore, we use $N_H=3\times10^{21}$\,cm$^{-2}$ as derived
from the CCD spectroscopy.

Figure~\ref{fig:soft_excess} shows the soft excess for each epoch
in comparison with other stars. T~Tau's 
soft excess is statistically equal in all
epochs, and greatly surpasses the excess seen in other 
young stars.

\begin{table*}
\begin{center}
%  \sidecaption
           \caption{Line ratios. Values in brackets are $1\,\sigma$ confidence ranges. \label{tab:grating_ratios}}
            \setlength{\tabcolsep}{0.1cm}
 \begin{tabular}{ l cccccccc }
  \hline
  \hline
   Line ratio  &\multicolumn{1}{c}{ All}                & \multicolumn{5}{c}{Date}  \\
                                        &   &   \multicolumn{1}{c}{2005~Aug~15th} & \multicolumn{1}{c}{2014~Aug~15th} &      \multicolumn{1}{c}{2014~Aug~25th} & 2014~Sep~6th & 2015~Jan~1st\\
    \hline 
     O {\sc vii} He$\alpha$ / He$\beta$ &  & 4.7 (3.2 -- 6.8) & 3.1 (2.2 -- 4.7) & 2.4 (1.7 -- 3.6) & 5.5 (3.8 -- 7.9) & --\\     
     O {\sc vii} r / O {\sc viii} & & 0.8 (0.6 -- 1.1) & 0.6 (0.5 -- 0.8) & 0.6 (0.4 -- 0.8) & 0.8 (0.6 -- 1.0) & -- \\
     O {\sc vii} G ratio (f+i)/r  & 0.59 (0.43 -- 0.85) & 0.97 (0.65 -- 1.36) & 1.16 (0.81 -- 1.75) & 1.16 (0.77 -- 1.72) & 0.535 (0.34 -- 0.81) & -- \\
     O {\sc vii} R ratio f/i  & >3.1 & >2.2 & >2.0 & 3.8 (>1.9) & >2.6 & --\\
     Ne~{\sc ix} G ratio\tablefootmark{a} & &&&&&1.3 (>0.8)\\
     Ne~{\sc ix} G ratio\tablefootmark{b} & &&&&& 0.9 (0.6--1.4)\\
     Ne~{\sc ix} R ratio\tablefootmark{a} & &&&&& >2.5 \\
     Ne~{\sc ix} R ratio\tablefootmark{b} & &&&&&2.2 (>1.2)\\
        \hline  
  \end{tabular}\\
  \tablefoot{
    \tablefoottext{a}{Positive MEG grating order only.}
    \tablefoottext{b}{Both MEG grating orders.}
  }  
\end{center}
\end{table*}

\subsection{Neon}
The Ne~{\sc ix} triplet (with wavelengths of 13.45, 13.55, and 13.70\,\AA{} for the recombination, intercombination, and forbbiden lines, respectively) can trace the density of plasma 
emitted at higher temperatures than O~{\sc vii} (approximately 4 versus 2\,MK), 
though with the caveat of a higher coronal contribution than in O~{\sc vii}.
The spectral resolution of the RGS is insufficient to reliably resolve
the triplet components from the ``contamination'' by nearby Fe lines so that we 
concentrate on the HETG.

Close inspection of the two MEG gratings orders reveals that only two photons are detected 
in the f+i lines of the negative grating order, 
which is statistically quite unlikely given the number of r-line photons (11) and
the photon number detected in the positive grating order at similar effective
area (11 in the f+i lines). This discrepancy might be related to an unknown (perhaps localized) detector artifact
not fully accounted for by the reduction software, possibly
caused by the proximity of the f-line (and to a lesser degree of the i-line) 
to the chip gap in the negative grating order although our data are insufficient
to demonstrate this conclusively. 
We focus on the positive grating order in the following, but also 
provide the results for the combined data for comparison in square brackets. Due 
to the low number of  counts, the confidence ranges usually overlap making this 
decision less critical for the physical interpretation.

The G-ratio above 0.8 [0.6] indicates a plasma temperature below 
$4.4\times10^6$\,K [$6.3\times10^6\,$K]. This points to an origin in 
the low-temperature plasma component -- similar to the oxygen {\sc vii} 
triplet. The R-ratio above 2.5 [1.2] indicates densities below 
$\log n_e = 11.4$ [$12.1$]. 
Factoring in the coronal contribution, the limits are 11.9 [13.3]. In summary,
the data indicate that the Ne~{\sc ix} emission originates in a rather cool (a few MK), low 
density plasma, although the formal 90\,\% confidence limit using both grating orders 
is not very restrictive.

\begin{figure}[t!]
\centering
\includegraphics[width=0.48\textwidth]{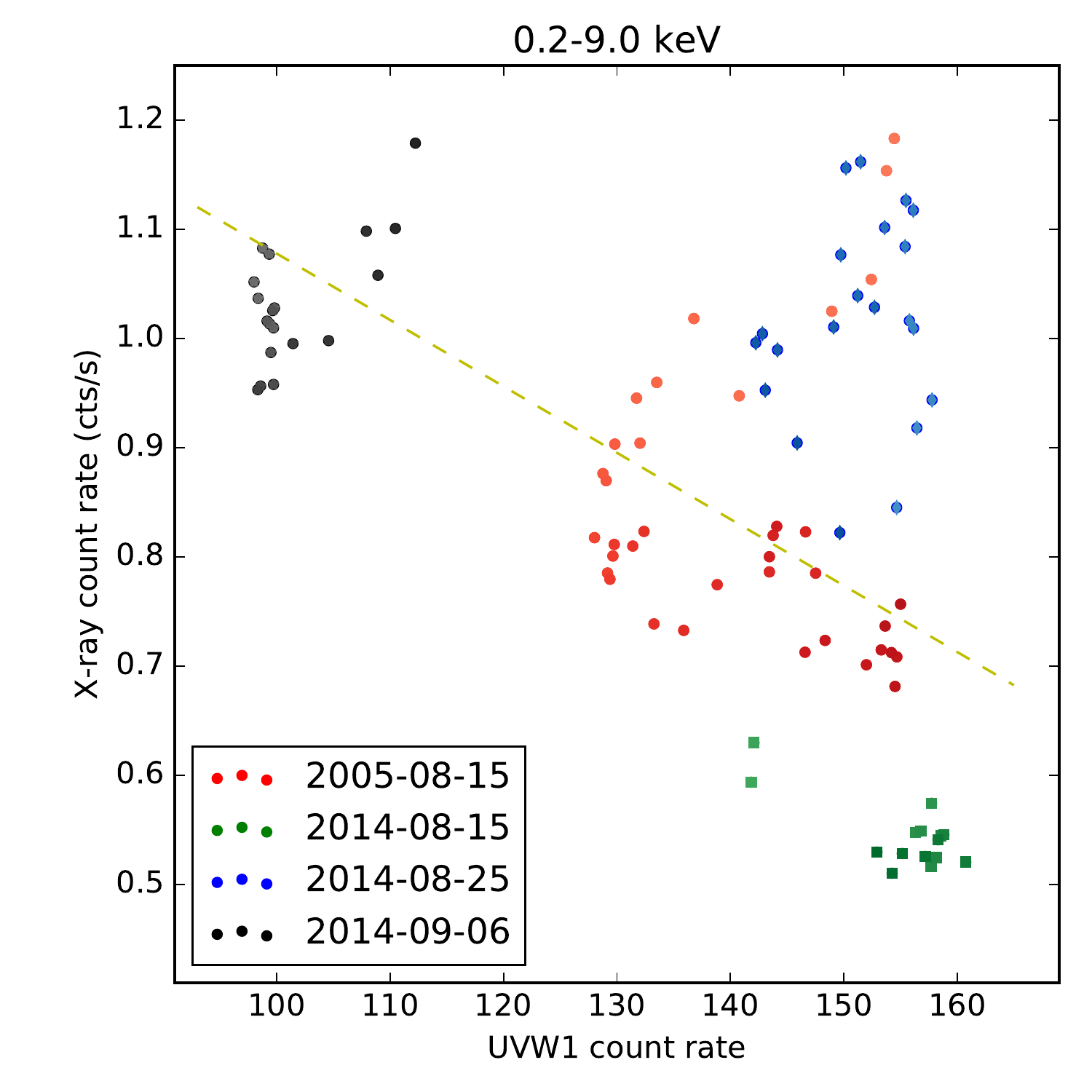}
\caption{X-ray count rates (pn) compared to OM UVW1 count rates with
colors indicating the individual epochs. OM data are averages within
the respective 2\,ks pn light curve bins.
\label{fig:OM_vs_X}}
\end{figure}

\begin{table*}[t]
           \caption{Exposure averaged UV count rates and fluxes, and 
           mean contemporaneous H$\alpha$ properties. H$\alpha$ errors are 
           estimated from the standard deviations from the observations falling within one day of
           the X-ray observation.\label{tab:UV_Fluxes}\label{tab:Ha}}
            \setlength{\tabcolsep}{0.1cm}
\centering            
 \begin{tabular}{ l c c c c c }
  \hline
  \hline
     Date & Count rate & NUV Flux $f$ & Dereddened NUV Flux & H$\alpha$ EW & H$\alpha$ 10\% width\\
          & (cts s$^{-1}$) & ($10^{-14}$ erg\,s$^{-1}$ cm$^{-2}$ \AA$^{-1}$) & $\log f$\\
     \hline
        2005-08-15 & 127 & 6.0 & -11.88 & \multicolumn{2}{c}{--}\\
        2014-08-15 & 168 & 8.0 & -11.76 & \phantom{ARAS: } $100\pm13$ & $595\pm10$\\
        2014-08-25 & 151 & 7.2 & -11.81 & \phantom{ARAS: }$97\pm2$ & $566\pm7$\\
        2014-09-06 & 100 & 4.8 & -11.99 & \phantom{ARAS: }$86\pm1$ & $439\pm2$\\
        2015-01-01 & \multicolumn{3}{c}{Not available (Chandra data)} & TIGRE:\phantom{$87$}--\phantom{12} & $570\pm58$ \\
                   &                          && & ARAS: $87\pm2$ & $566\pm10$\\
        \hline
  \end{tabular}                           
\end{table*}

\section{X-ray versus NUV emission \label{sect:XvsUV}}
Previous studies provided evidence for a negative 
correlation between X-ray luminosity and accretion rate
in different samples of stars and non-simultaneous data
\citep[e.g.,][]{Telleschi_2007}. We use the T~Tau data to check 
this correlation on a single star using pn and UVW1 count
rates as tracers of 
X-ray luminosity and
accretion rate \citep[the UVW1 band is dominated by accretion shock emission; see][]{Calvet_2004}. 
Stellar activity contributes to both wavelength regimes, which likely 
weakens any signal from correlated accretion-X-ray variability. However,
only two short events show correlated X-ray-UV variability that 
might be related to flares so that most of the variability must have
a different origin.

Figure~\ref{fig:OM_vs_X} shows the correlation between X-ray 
emission and NUV count rate.
% We specifically plot the hard 
% (1.0 -- 5.0\,keV) and soft (0.2 -- 0.6\,keV) energy bands, because 
% the hard band is largely dominated by emission from coronal activity 
% while any accretion generated photons are expected in the soft band.
% 
% 
There is a tendency for the X-ray count rate to decrease
with increasing NUV flux. Taking the XMM-Newton energy 
band (0.2 -- 9.0\,keV), we find 
a significant negative correlation (--0.39 with a p value of $10^{-4}$ for
Spearman's R; similar results are obtained for Kendall's $\tau$ and Pearson~R).
We examined individual energy bands and find that the correlation is driven by the energy
band providing the largest count numbers while 
there is no significant change
of the spectral hardness as a function of UV 
count rate. It is noticeable, however, 
that the soft band ($E_{phot} < 0.6\,$keV) 
shows no significant correlation with UV count rate.

% , we find that the most significant correlations
% pertain to energy bands with a lower bound between 0.6 and 1.0\,keV
% and an upper bound of 1.5\,keV 
% (correlation coefficients around --0.5 for Spearman~R and Pearson~R (--0.36 for Kendall's $\tau$)
% and p~values $<1\times10^{-6}$ for all).  
% For both, higher and lower photon energies, the correlation 
% is flatter and the significance is weaker. We also investigated ,
% but found no significant correlations, which shows that the X-ray flux 
% uniformly scales with UV count rate.

At first sight, the correlation appears to be driven by the 2014~Sep~6th
observation. Omitting
this particular epoch still results in negative 
albeit less significant correlations. 
On the other hand, only some epochs show negative 
correlations while others show positive correlations.
Therefore, the overall anti-correlation only applies on average, but not on 
short timescales.

To compare our results with published relations, which are typically 
expressed as power laws
($\log L_X = a + b\log \dot{M}$), we approximate the  conversion from 
count rates to X-ray/accretion 
luminosities by first order polynomials (e.g., 
accurate to within 6\,\% for X-ray luminosities) and ignore any photospheric
contribution to the UVW1 flux (accurate to about 10\,\%). 

\begin{table*}[t]
\begin{center}
           \caption{Comparison of T~Tau with some other CTTSs and HAe stars
           that have high-resolution X-ray grating spectroscopy. \label{tab:comparison}}
            \setlength{\tabcolsep}{0.1cm}
 \begin{tabular}{ l cccccrcccc }
  \hline
  \hline\\[-0.3cm]  
   System & $M_\star$ & $R_\star$ & Rotation & $R_{co}$\tablefootmark{a} & $\overline{B}$ & typical $\dot{M}$ & Spot covering & Inclination & Refs.\\
          & ($M_\odot$) & ($R_\odot$) & Period (d) & (au/$R_\star$) & (kG) & ($M_\odot$\,yr$^{-1}$) & fraction (\%) & ($^\circ$)\\
   \hline
        TW~Hya & 0.7 & 0.8 & 4.7 & 0.05/13 & 2.6 & a few $10^{-9}$ & 1.1 & $\sim$ face-on &  10, 4, 5 \\
        RU~Lup & 0.8 & 1.7 & 3.7 (?) & 0.04/5.5 & (perhaps 3) &  $4\times10^{-8}$ & & & 7, 8, 10\\
        BP~Tau & 0.8 & 2.1 & $\approx8.3$ & 0.07/7.6 & 2.1-2.2& $2.9\times10^{-8}$ &  2.2 & &  14, 4, 13, 10, 15 \\
        MP~Mus & 1.2 & 1.3 & 5.8 & 0.07/11 & -- &  $5\times10^{-9}$ &  45 & 24, 25\\
        T~Tau & 2.2 & 3.6 & 2.5 & 0.05/3 & 2.6 & $10^{-7}$ & 3.3 & $\sim$ face-on & 1, 2, 3, 24\\        
        HD~163296 & 2.2 & 2.3 & (2.1?) & 0.02/1.6 & $<0.1$ & a few $10^{-7}$  &1.8 & 38 & 16, 22\\ 
        AB~Aur & 2.5& 2.6 & ($\approx2$?)& 0.04/3.5  & $<0.2$& $1.4\times10^{-7}$&  & $40$ & 18, 19, 20, 22\\
        \hline
  \end{tabular}   
  \end{center}
  \tablefoot{
  \tablefoottext{a}{From stellar rotation period assuming Keplerian rotation.}
  }
%   \caption

 \tablebib{
  (1) \citet{Guedel_2007}; (2) \cite{Podio_2014}; (3) \citet{Bouvier_1995}; (4) \citet{Johns-Krull_2007}; (5) \citet{Dupree_2012}; (7) \citet{Lamzin_1996}; (8) \citet{Stempels_2002}; (9) \citet{Herczeg_2005} and references therein; (10) \citet{Brickhouse_2010} and references therein; (11) \citet{Siwak_2016}; (12) \citet{Johns-Krull_1999}; (13) \citet{Xiao_2012}; (14) \citet{Furlan_2011};
  (15) \citet{Ingleby_2013}; (16) \citet{Mendigutia_2012}; (17) \citet{Gregorio_2013}; (18) \citet{Garcia_2006}; (19) \citet{Cauley_2015} and references therein; (20) \citet{Fairlamb_2015}; (21) \citet{Alecian_2013}; (22) \citet{Deleuil_2005}; (23) \citet{Calvet_2004}; (24) \citet{Batalha_1998}; (25) \citet{Pascucci_2007}.}
\end{table*}
% Co-rotation radius ((Prot/365.25)^2 * Mstar)^(1/3)

With these transformations, we find that the X-ray luminosity scales as
$\log\,L_X \sim - (0.74\pm0.13) \log \dot{M}_{acc}$, where we considered
the error in accretion and X-ray luminosity using total least squares. Specifically, to 
compare our results with \citet{Telleschi_2007},  
we consider the so-called residual 
X-ray luminosity, that is, we 
divide the X-ray luminosity 
by the expected X-ray luminosity based on stellar mass\footnote{This expected 
X-ray luminosity can be based on stellar mass or on the relation between 
stellar mass and mass accretion rate, $\log \dot{M} = 2\log M - 7.5$.} 
($\log L_X^{expected} = 30.88$) and find
$L_X^{res} = -5.1^{+0.5} - (0.74\pm0.13) \log \dot{M}_{acc}$
with an unconstrained lower bound for the constant term due to the interplay with the 
linear term. This relation is compatible with the \citet{Telleschi_2007} power law found 
for a sample of stars with non-simultaneous accretion rate measurements
($\log\,L_X/L(\dot{M}) =-(4.05\pm1.19) - (0.45\pm0.15) \log \dot{M}_{acc}$).
Therefore, the residual X-ray luminosity of T~Tau varies with accretion rate in
the same sense as observed for samples of 
T~Tauri stars.

\section{Discussion \label{sect:disc}}
% \color{blue}
\subsection{Comparison of T~Tau's X-ray grating spectroscopy to other CTTSs and HAe stars}
We first concentrate on the line diagnostics from our X-ray grating spectroscopy. The O~{\sc vii} (and to a lesser degree the Ne~{\sc ix}) 
line ratios differentiate T~Tau from other, lower mass CTTSs, which generally show 
high densities in the O~{\sc vii} and Ne~{\sc ix} triplets. 
In particular, we measure an O~{\sc vii} R-ratio of $\gtrsim2$ for T~Tau, 
while other single CTTSs have R-ratios between 0.2 and 0.4; for example $0.21\pm0.07$ \citep[TW~Hya,][]{Brickhouse_2010},
$0.3\pm0.2$ \citep[RU~Lup,][]{Robrade_2007}, $0.37\pm0.16$ \citep[BP~Tau,][]{Schmitt_2005}, or $0.28\pm0.13$
\citep[MP~Mus,][]{Argiroffi_2007}. A similar pattern is found in 
Ne~{\sc ix}, where our lower limit 
is below the value derived for TW~Hya \citep[$>2.5$ in T~Tau versus $0.51\pm0.03$ 
in TW~Hya,][]{Brickhouse_2010}. 

T~Tau's oxygen and neon R-ratios are more comparable to those found in HAe~stars, where  
X-ray grating spectra mostly indicate low densities 
\citep[e.g., AB~Aur, HD~163296, see][resp.]{Telleschi_AB_Aur, Guenther_2009}.
We will therefore use these as representative
for HAe X-ray data in the following. The HETG spectrum of 
the Herbig~Ae star HD~104237 shows densities around 
$10^{12}\,$cm$^{-3}$ \citep{Testa_2008} and does not fit this pattern.
However, HD~104237 is a spectroscopic
binary, which makes conclusions based on this measurement less robust than for the 
two bona~fide single HAe~stars. 

The similarity of T~Tau with HAe stars in terms of X-ray density diagnostics
suggests that the mass accretion rate, which T~Tau shares with HAe stars
(e.g.,\, $10^{-7}\,M_\odot\,$yr$^{-1}$ for HD~163296 while TW~Hya has only a few
 $10^{-9}\,M_\odot\,$yr$^{-1}$), is more important than the
magnetic field strength, which T~Tau shares with CTTSs but not with
HAe stars \citep[e.g., the upper limit for HD~163296 
is about 50\,G; see][]{Hubrig_2007}. 
Therefore, we concentrate on the properties of the accretion shock
to explain the differences between T~Tau, CTTSs, and HAeBe stars, and ignore,
for the moment, the effect of the magnetic field, which is
clearly a strong simplification.
In fact, there is evidence from Zeeman-Doppler Imaging
that magnetic fields on higher-mass CTTSs are likely
more complex than on their lower-mass siblings, which have predominately dipolar magnetic
fields \citep{Gregory_2012}. This probably also influences the accretion process 
and causes 
higher accretion shock densities for magnetic fields dominated by higher multipole moments 
\citep{Adams_2012}. Nevertheless, we argue
that the shock itself remains largely unaffected as long as the magnetic field
strength is sufficient to channel the material onto the stellar surface.
Furthermore, the similarity in the X-ray-derived densities between T~Tau and, for example, the HAe star HD~163296,
despite their strongly contrasting magnetic field configuration, argues against
a decisive impact of the magnetic field on the observed X-ray emission from
the accretion shock.

\subsection{The origin of the O~{\sc vii} emission and the soft excess}
We now argue that the O~{\sc vii} emitting plasma cannot be associated
with direct post-shock emission. 
We specifically consider three example cases: one representative 
of the well-studied low-mass, magnetic CTTS (TW~Hya);
one representative of the non-magnetic HAe stars (HD~163296); and, T~Tau as being squarely intermediate between these two examples. Their general properties as well as those for other young, accreting 
stars with X-ray grating spectroscopy are summarized in Table~\ref{tab:comparison}. 

The accretion rates derived based on X-ray fluxes generally tend to fall short 
of those from more traditional accretion tracers \citep[e.g., from optical emission lines fluxes, see][]{Curran_2011}.
In particular, the O~{\sc vii} density diagnostics show that the plasma cannot carry the bulk of the accretion rate,
simply because the total accretion rate would be too low for T~Tau (see Sect.~\ref{sect:O7-R}) and HD~163296 
\citep[see][]{Guenther_2009}. A comparable argument applies to TW~Hya: \citet{Brickhouse_2010} locate the O~{\sc vii} emitting plasma
in the low-density ``splatter'', because it is, first, subject to a lower absorbing 
column density than the Ne~{\sc ix} emitting plasma (which is likely associated
with the accretion shock) and, second, has a lower density than the 
Ne~{\sc ix}  emitting plasma, which is opposite to expectations from 
shock models. In other words, the mass-accretion rates associated with the O~{\sc vii} emitting plasma fall short of the rates derived from other accretion tracers by almost one (TW~Hya) to several orders of magnitude (T~Tau and HD~163296),
when using literature values for the accretion spot sizes, the constraints on the plasma density, and reasonable values for the infall velocity (see references provided in Table~\ref{tab:params}).

In addition, requiring that the O~{\sc vii} emitting plasma has a similar spatial footpoint 
to the accretion spot(s) provides an insufficient  amount of
post-shock emission measure (EM) for T~Tau and HD~163296 (see Table~\ref{tab:params}).
%  provides insufficient 
% emission measure as can be seen from the following estimates. We use the fiducial values provided 
% in Tab.~\ref{tab:params}
Specifically, we approximate the radiative cooling length as
\begin{equation}
d_{cool} = 3\times10^{6}\,\text{cm} \left(\frac{10^{10}\,\text{cm}^{-3}}{n_e} \right) 
\left(\frac{\textrm{v}_{shock}}{100\,\text{km}\,\text{s}^{-1}}\right)^{4.5} \label{eq:dcool}
,\end{equation}
derived from Fig.~5 in \citet[][with the infall (or shock) velocity v$_{shock}$
and the pre-shock electron density  $n_e$, which is related to the
measured (=post-shock) density via $n_e = n_{measured}/4$]{Guenther_2007}. The upper limit\footnote{This simplified treatment assumes constant, high temperature within $d_{cool}$ while the temperature gradually decreases with increasing distance from the shock location in reality.} for the 
emission measure of the post-shock plasma
is then
\begin{align*}
EM_{sh}       & =  n_e^2 V = n_e^2 d_{cool} A_{shock}   \\
        & =  3\times10^{46} \text{cm}^{-3} \left(\frac{n_e}{10^{10}\,\text{cm}^{-3}}
        \right) \left(\frac{\textrm{v}_{shock}}{100\,\text{km}\,\text{s}^{-1}}\right)^{4.5}
    \left(\frac{A_{shock}}{10^{20}\,\text{cm}^2} \right)\,, \nonumber
\end{align*}
% or
% \begin{align*}
% A_{shock} & =  EM_{sh} n_e^{-2} d_{cool}^{-1}     \\
%  & = 3.3\times10^{26}\,\text{cm}^2 \left(\frac{EM_{sh}}{10^{53}\text{cm}^{-3}}\right)
%          \left(\frac{10^{10}\,\text{cm}^{-3}}{n_e}
%         \right) \left(\frac{100\,\text{km}\,\text{s}^{-1}}{\textrm{v}_{shock}}\right)^{4.5}
%     , \nonumber
% \end{align*}
where $A_{shock}$ is the surface area of the shock and we use the O~{\sc vii} measured densities for 
the post-shock density. We further use the $EM$ of the coolest plasma component,
in which the vast majority of the O~{\sc vii} emission originates, 
factoring out differences in oxygen 
abundance\footnote{We multiply $EM$ with the oxygen abundance relative to T~Tau since 
the absolute metallicity is not well defined. Therefore, we consider 
these values more suitable for this comparison, but note that our 
conclusions are qualitatively insensitive to these details.} to avoid biasing this comparison. 
With this approach, we ignore any contribution from the corona to the
soft plasma component. In fact, the soft excess tells us that the corona
most likely contributes only a little (20-30\,\%) to the soft plasma component
assuming that coronae on CTTSs are scaled-up versions of
main-sequence star coronae.
%
% 
% Further assuming that the accretion shock does 
% not cover more than 10\,\% of the stellar surface (a rather generous upper limit), we find that 
% the post-shock emission measure falls short of the measured values by factor of three (for TW~Hya using the nominal values)
% up to almost two orders of magnitude (for T~Tau).
Clearly, the O~{\sc vii} emission in T~Tau and HD~163296 originates in a structure, which has a larger vertical height than the cooling distance while this is not necessary for TW~Hya although a more detailed treatment will likely reveal that 
the emissivity of the post-shock plasma is insufficient for the observed emission measure even in TW~Hya.
This is in line with \citet{Brickhouse_2010} who ``arbitrarily'' increase the vertical
scale height of the cool plasma to 0.1\,$R_\star$ in order to produce the required emission measure.
% as they assume a higher pre-shock density derived from the Ne~{\sc ix} triplet\footnote{For TW~Hya, 0.1\,$R_\star$ corresponds to about ca. 150 times the cooling distance as they assume a pre-shock density of $\log n_e = 11.8$ resulting in a cooling length of 400\,km; Eq. (1) gives 700\,km for their values}.

In summary, we find that (a) only a small fraction of the O~{\sc vii} emission can be potentially associated with post-shock cooling plasma and (b) the O~{\sc vii} emission can only represent a small fraction of the accretion luminosity. Therefore, we need a mechanism that increases the 
scale height of the cool plasma while simultaneously blocking most of the direct post-shock emission.

\subsection{Photo-heated, density stratified accretion columns \label{sect:model}}
\begin{figure}[t!]
\centering
\includegraphics[width=0.48\textwidth, trim=5mm 10mm 5mm 20mm, clip]{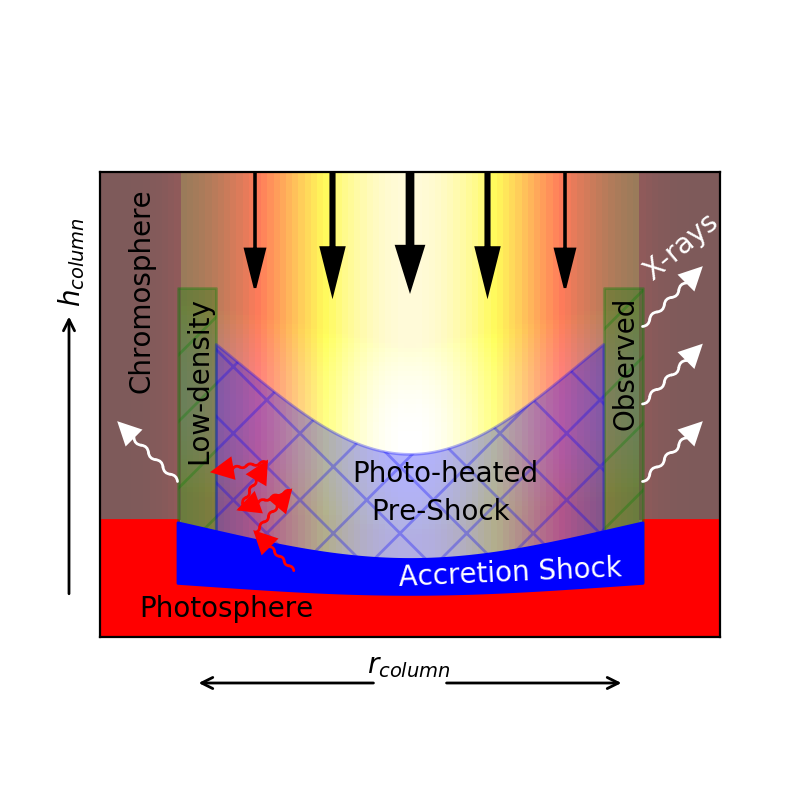}
\caption{Not-to-scale sketch of the envisaged accretion shock geometry. The observed X-rays originate in the outer shell of the accretion stream, which carries only a small fraction of the total accretion rate.
\label{fig:sketch}}
\end{figure}

We propose radially density  
stratified accretion columns (see sketch in Fig.~\ref{fig:sketch}).
This model is based on the 
global Magneto-Hydrodynamic (MHD) models of \citet{Romanova_2004} and includes the radiation transport studied by 
\citet{Costa_2017}. First, the MHD models show that accretion 
columns have a dense
inner core surrounded by lower density material, while, second, the
radiation from the hot post-shock plasma heats 
the pre-shock material and 
increases the scale height of the hot plasma. 
We envision the density profile of the stream to increase radially 
from the outside to inside, which determines the 
optical depth to each point of the X-ray emitting post-shock plasma
and, thus, also the mean density of 
the plasma observed at X-ray wavelengths and the fraction of the 
energy released in the shock that escapes at X-ray wavelengths.

The radial density stratification results, first, in 
a dense inner core of the accretion stream that 
provides the bulk of the accretion material,
but is largely invisible
in X-rays due to absorption by the  incoming (pre-shock) gas and, second, 
increases the vertical height of the hot plasma since the post-shock 
emission is not only absorbed by the pre-shock material at the same 
streamline \citep[as in the model by][]{Costa_2017} but also cannot 
escape unhindered radially as the aspect-ratio (height/radius) is around 
unity for T~Tau and HD~163296 (2.4 and 0.2, respectively). For TW~Hya, the aspect ratio is 
$0.01$ implying that the post-shock radiation can escape more easily in 
the vertical direction assuming some moderate tilt in the accretion column.
The observed O~{\sc vii} emission does not come entirely 
from the post-shock plasma. In our scenario, the pre-shock plasma 
contributes, too. The high-energy photons from the inner, high-density
regions of the accretion shock are mostly re-processed in the pre-shock 
material, both on the same and on other streamlines. These photons heat
and ionize part of the infalling material before it goes through the terminal 
shock at the stellar surface. Photons from the low-density, outer parts of the 
accretion stream dominate the emission from the pre-heated material, because
these regions suffer less extinction toward the observer and have longer 
equilibrium timescales than the dense, inner accretion stream regions.
Re-processing only a
small fraction of the energy released in the accretion-shock into O~{\sc vii}
emission is sufficient to cause the soft excess.
In addition to the high-energy post-shock photons, a 
reverse shock may also contribute to the pre-heating of the accretion funnels
\citep[see laboratory experiments by][]{Revet_2017}. In reality, some 
combination of these two processes is likely responsible for the 
O~{\sc vii} emitting plasma in the low-density regions of \citep[or around,][]{Orlando_2013} 
the accretion funnel above the stellar surface since other mechanisms 
are thought to be insufficient \citep[such as thermal conduction, see][]{Orlando_2010, Revet_2017}.
In order to thoroughly test this scenario, one needs detailed MHD-radiation transfer calculations,
which are beyond the scope of this paper. Also, the above scenario
of radially stratified accretion columns does not exclude other
contributions to the soft excess such as a splatter or a reverse shock in the case of high-$\beta$
conditions\footnote{The value $\beta$  describes
the ratio between plasma and magnetic pressure.} in the post-shock region \citep{Brickhouse_2010, Revet_2017}.

% Since the pre-shock material is rather dense, equilibrium time scales 
% are well below the flow times \citep{Smith_2010}. We envision that the
% pre-heating skews the observed emission measure distribution
% slightly lower temperature with respect of the immediate post-shock emission.

Stars like TW~Hya with a low accretion rate also have a low pre-shock column density
that absorbs a comparably small fraction of the immediate X-ray emission 
from the accretion shock and the observed emission is dominated by the dense 
inner part of the accretion stream. With increasing mass accretion
rate (T~Tau and HD~163296), a larger fraction of the post-shock
emission must pass through large columns. Thus, this emission does not escape,
but heats the pre-shock plasma above the accretion shock. This
increases the aspect ratio and decreases the observed density as
only the emission originating in the outer, low density shells can
escape relatively unhindered.

\begin{table}[t!]
\begin{center}
           \caption{Physical parameters used. References are provided in square brackets. \label{tab:params}}
            \setlength{\tabcolsep}{0.1cm}
   \begin{tabular}{ l ccc}
  \hline
  \hline\\[-0.3cm]  
   Parameter         & T Tau & TW Hya & HD 163296 \\
  \hline\\[-0.3cm]  
  normalized EM ($10^{52}\,$cm$^{-3}$)  & 50 [1] & 7 [2] & 4.4 [3]\\
  Spot covering fraction (\%) & 3 [4] & 3 [5] & 8 [6] \\
%   Spot area ($20^{22}$\,cm$^2$) & 2.4 & 0.12 & 2.6\\
  Spot radius ($10^{10}$\,cm) & 9 & 2 & 9 \\
  $\log$ X-ray density\tablefootmark{a}  & <10.4 & 11.8 & <10.4 \\
  $\log$ pre-shock density\tablefootmark{b} & 9.5 & 11.3 & 9.5\\
  $v_{shock}$ (km/s) & 400 & 500 & 450\\
  $d_{cool}$ ($10^8$\,cm) & $49$ & $2.1$ & $83$\\
  Vertical height ($10^{8}$\,cm) & $2.1\times10^3$ & $1.5$ & $171$\\
  Associated $\dot{M}_{acc}$ ($10^{-10}\,M_\odot\,\textrm{yr}^{-1}$) & 1.1 & 4.3 & 1.4\\    
  Typical $\dot{M}_{acc}$ ($10^{-9}\,M_\odot\,\textrm{yr}^{-1}$) & 100 & a few & a few 100\\
  \hline
  \end{tabular}
  
   \tablefoot{
  \tablefoottext{a}{Measured from the O~{\sc vii} triplet.}
  \tablefoottext{b}{Based on the post-shock density from the O~{\sc vii} triplet.}
  }
%   \caption

 \tablebib{
  [1] This work; [2] \citet{Brickhouse_2010}; [3] \citet{Guenther_2009}; 
  [4] \citet{Calvet_2004}; [5] \citet{Batalha_2002}; [6] \citet{Mendigutia_2013}.
  }
\end{center}  
\end{table}

For a fiducial area of the dense,
inner stream of one tenth of the outer, low density part, that is, 
covering fractions of 0.3\,\%, 0.3\,\%, and 0.8\,\% for T~Tau, TW~Hya, 
and HD~163296, respectively, we find that densities of
$\log n=13.5$, $13.1$, $13.8$ for T~Tau, TW~Hya, and HD~163296, 
respectively are required in the inner stream to carry the accretion 
rates derived from established accretion
diagnostics (e.g., UV excess). 
These high densities result, first, in rather short cooling distances so that the 
radius of the stream largely exceeds the shock cooling distance (cf. Fig.~\ref{fig:sketch}). 
Second, this inner, high density shock  results in large emission 
measures associated with the core of the accretion stream. 
The emission measures associated with the post-shock cooling zone of the inner funnel 
would outshine the outer, low density part of the accretion stream (and the corona) 
without absorption for T~Tau and HD~163296 ($EM> 10^{54}$\,cm$^{-3}$). 
The emission from this dense plasma is, however, absorbed in the pre-shock plasma. For example, 
10\,\% of the inner, dense column in the radial direction already corresponds to a 
column density of $>2\times10^{22}$\,cm$^{-2}$, which extinguishes 
the O~{\sc vii} and Ne~{\sc ix} triplets in T~Tau and HD~163296.
Since the fraction
of the soft X-ray to accretion emission is largest in TW~Hya, the emission 
measure associated with 
the inner part of the accretion spot is comparable to the one observed ($EM=7\times10^{52}$\,cm$^{-3}$) 
as expected.
% (the "self-absorbing column" is only a tenth of the values for T~Tau and HD~163296, which
% is sufficiently less due to the exponential nature of the absorption). 
While the innermost part
of the stream remains obscured, the emission from the slightly less dense surrounding shells 
can escape, thus explaining the high densities.

The densities of the inner stream that we derive here exceed the 
typically quoted values 
\citep[$\sim10^{13}$\,cm$^{-3}$, see review by][]{Hartmann_2016}. These densities
are derived from traditional accretion tracers such as UV excess
assuming a homogeneous density in the accretion stream. However, the 
average density of the radially density stratified accretion 
stream is lower than the
values for the inner stream so that the densities in our scenario compare
quite well with the literature values.
 
The deep X-ray grating spectrum of TW~Hya allows us to perform an additional 
test of our scenario. The excess absorption provided by the O~{\sc vii} emitting region, 
that is, the ``splatter'' in the \citet{Brickhouse_2010} model, is
the low density outer stream in our model.
The absorbing column density
towards the O~{\sc vii} emitting region is $2\times10^{20}$\,cm$^{-2}$
while the column density
towards the Ne~{\sc ix} emitting region is $2\times10^{21}$\,cm$^{-2}$.
Assuming a linear increase in density from the O~{\sc vii} emitting
region ($n_e = 6\times10^{11}\,$cm$^{-3}$) to the Ne~{\sc ix} emitting 
region ($n_e=3\times10^{12}$\,cm$^{-3}$),
the required distance to accumulate an absorbing column density
of $1.8\times10^{21}$\,cm$^{-2}$ is $10^9$\,cm, which is about
the value estimated by \citet{Brickhouse_2012} for a constant density.
Given the 
simplicity of this model, this value is surprisingly close to 
the sizes estimated above for the inner, high density part of the 
accretion stream in TW~Hya ($2\times10^9$\,cm).

This scenario also mimics individual aspects of models already proposed in the 
literature. From an X-ray point of view, it includes the absorption effects
proposed by \citet{Brickhouse_2010} as the outer layer of the accretion
stream has a similar role as the ``splatter'' in the \citet{Brickhouse_2010}
model, that is,  it is responsible for 
the softest X-ray photons while also absorbing the central region
of the accretion shock. It includes
the optical depth effects discussed by \citet{Sacco_2010} and \citet{Curran_2011}
as the cause of the low-mass accretion rates derived
from X-ray (O~{\sc vii}) line fluxes. In our model, however, the 
observed X-ray emission is not caused by ``appropriate'' streams, but
rather by the ``appropriate'' outer shells of the stream.  
Also, our model includes optical depths effects already discussed by 
\citet[][mainly for resonance line scattering, which would massively affect
the G-ratio for which we do not find any hint in our data; the observed G-ratio is 
fully compatible with those expected for a thermal plasma emitting 
O~{\sc vii} photons]{Argiroffi_2009} and in more detail by \citet[][]{Bonito_2014}.

 \color{black}
\subsection{Accretion versus X-ray emission}
Analysing the EPIC~pn and OM data, we find 
that higher UV fluxes correspond, on average, to
lower X-ray luminosities. We interpret this as an anti-correlation
between accretion and (magnetic) X-ray activity, because
the majority of the observed X-ray and UV variability is rather 
smooth and happens on timescales
comparable to the duration of the observations (half a day and 
longer). Flaring activity could affect this relation, but
only two short-duration, weak events are present so that 
we regard stellar activity as unlikely to dominate over accretion effects
in our data. Similarly, 
time lags between accretion-driven UV excess and X-ray response 
\citep[e.g.,][]{Dupree_2012} or vice versa
might weaken the correlation but are not expected to spoil the correlation 
given the smooth variations
(we also checked that reasonable time lags do not improve the correlation).
% Therefore, we proceed by discussing the observed correlation in terms
% of correlated accretion-X-ray luminosity variability.

Our analysis of the accretion-X-ray 
luminosity correlation 
is complementary to published
studies, which rely on (mostly) non-simultaneous 
data  for different samples of stars
\citep[e.g.,][]{Neuhaeuser_1995,Preibisch_2005, Telleschi_2007, 
Guedel_2007_XEST, Bustamante_2016} while we investigate the correlation
between accretion and X-ray luminosity for a single star. Interestingly,
our correlation is compatible with published studies, for example,  \citet{Telleschi_2007} 
for the XMM-Newton Extended Survey of Taurus (XEST) sample \citep{Guedel_2007_XEST} and 
\citet{Bustamante_2016} who estimate accretion rates based on U-band excess
for Orion Nebular Cloud (ONC) stars, a method that is similar to our UV excess approach.

Given that the correlation between accretion 
rate and X-ray luminosity
observed for samples of stars extends to different accretion rates 
for an individual star, we shall interpret these correlations either
in terms of accretion suppressing (detectable) X-ray emission (a causal
effect) or in terms of a third property that influences both X-ray 
emission and accretion. Other phenomena, for example, stellar evolution, can 
be excluded. 

On an individual star, accretion variability on day to week timescales is thought to be mainly due to rotational modulation; only 
rare accretion bursts occasionally enhance the accretion rate
\citep[e.g.,][]{Costigan_2012, Costigan_2014, Venuti_2014,Rigon_2017}.
Following this reasoning, the variability in the accretion rate as
traced by the NUV flux is driven by T~Tau's rotation. The anti-correlation
between accretion rate and X-ray flux then implies that rotation also 
is responsible for part of the X-ray variability \citep[][already suggest that 
some of the X-ray variability is due to rotation]{Guedel_2007}. This might be explained by 
the ratio between closed and open 
field lines. The open field lines connect the star with the disk and carry 
the accretion material. For the Sun, we know that the X-ray emission originates 
in closed loops while open field lines contain thin plasma producing little 
X-ray emission. Therefore, the hemisphere covered with a larger fraction of 
open field lines shows stronger accretion signatures and emits less X-ray emission.

This would fit into our scenario of density stratified accretion columns.
The thin, outermost regions of the stream are unlikely to be visible in the X-ray regime 
or in other tracers, because of their 
low density, but might still interfere  with coronal structures thus reducing the harder X-ray emission.
This requires a very low density layer around the funnels, which covers 
a relatively large fraction of the star (a few 10\,\%).
This layer would not cause other observable features due to its low density;
especially the photosphere would 
remain largely unaffected by the impact of very low density material. Also, virtually
no extinction would be associated with the low density outermost shells of the 
accretion stream (for a fiducial value of $n=10^{9}$\,cm$^{-3}$, 
the absorbing column density for a length corresponding to the
stellar radius would be in the 
$10^{20}$\,cm$^{-2}$ range, which is low compared to the 
measured $N_H$ values).

\section{Summary and conclusions \label{sect:summary}}
We have analyzed four new X-ray observations of T~Tau and accompanying 
NUV data, H$\alpha$ profiles, and optical photometry. Our main observational results 
can be summarized as follows:
\begin{itemize}
  \item The strong X-ray soft excess is present in all four XMM-Newton observations
          spanning timescales from weeks to years.
  \item The O~{\sc vii} triplet is compatible
          with the low density limit in all four epochs. This demonstrates 
          that this feature is stable and that variability in T~Tau's accretion
          properties cannot explain the difference to its lower mass siblings.
  \item Similar to the O~{\sc vii} triplet, the Ne~{\sc ix} triplet shows low 
          densities. This suggests that the emission from plasma with 
          temperatures of a few MK is dominated by low density material.
  \item Accretion rates between $4$ and $8\times10^{-8}\,M_\odot$\,yr$^{-1}$ apply for the 
        times of the X-ray observations.        
  \item There is about a factor of two variability during and between the exposures 
        in all fluxes (soft, X-rays, hard X-rays, and NUV photometry). Most of the variability 
        happens smoothly on timescales comparable to the on target times. A few short, 
        isolated events exist but do not dominate the variability.
  \item X-ray and NUV variability is anti-correlated. This anti-correlation mirrors
        the anti-correlation between X-ray luminosity and accretion found in samples of 
        CTTSs.
\end{itemize}
We further discussed the low density finding and the soft excess in terms 
of radially stratified accretion columns and the X-ray-NUV flux anti-correlation
in terms of magnetic field effects. 

First, density stratified accretion 
funnels provide a natural explanation of the X-ray diagnostics of CTTSs
by merging features of existing models: accretion generated X-ray emission is only seen
from the outer, low density radii of the accretion funnel while the inner, high density
part of the 
stream is obscured from view but carries the majority of the accretion.
In this scenario, the origin of the soft excess is similar 
in all accreting stars while the difference in density diagnostics between low- and 
intermediate-mass stars is ascribed to different escape probabilities between the 
dense inner and thin outer parts of the accretion 
streams.  Higher accretion rates with more
important higher magnetic multipole moments in intermediate-mass stars result in lower observed
densities than the lower accretion rates and more important dipole components in lower-mass stars.

Second, the anti-correlation between X-ray flux and accretion rate found for T~Tau 
extends similar relations for samples of stars, which strongly suggests that
this anti-correlation is driven by a phenomenon intrinsic to processes on 
accreting stars. We suggest that the 
stellar magnetic field plays a key role through enhancing X-ray emission in
surface patches with closed magnetic field lines while simultaneously enhancing
accretion signatures in patches with large fractions of open field lines.

In the future, a more systematic study of the transition region between 
high and low density X-ray diagnostics in CTTSs and HAe stars can help to 
confirm the proposed model of density stratified accretion funnels, which would
be ideally supplemented by magnetic field studies on the very same stars. In addition,
the XMM-Newton archive contains numerous observations of CTTSs during which the OM was
operated with UV filters that include accretion driven emission. Hence, extending
our correlation analysis between UV and X-ray count rates can extend the short term 
correlation between accretion rate and X-ray luminosity on individual stars. A broader
data coverage would solidify the accretion-X-ray anti-correlation.

\begin{acknowledgements}
PCS gratefully acknowledges support by DLR~50~OR~1307 and an ESA Research Fellowship 
during which part of the analysis was conducted. HMG acknowledges support by NASA-HST-GO-12315.01.
The results reported in this article are based on observations made by the Chandra X-ray and the XMM-Newtons
Observatories. 
This work also made use of data from the European Space Agency (ESA)
mission {\it Gaia} (\url{https://www.cosmos.esa.int/gaia}), processed by
the {\it Gaia} Data Processing and Analysis Consortium (DPAC,
\url{https://www.cosmos.esa.int/web/gaia/dpac/consortium}). Funding
for the DPAC has been provided by national institutions, in particular
the institutions participating in the {\it Gaia} Multilateral Agreement.
Support for this work was provided by the National Aeronautics and
Space Administration through Chandra Award Number GO5-16014X issued by
the Chandra X-ray Observatory Center, which is operated by the
Smithsonian Astrophysical Observatory for and on behalf of the National
Aeronautics Space Administration under contract NAS8-03060.
\end{acknowledgements}

\bibliographystyle{aa}
\bibliography{ttau}

\appendix

\section{Absorption from O~{\sc vii} He-like lines \label{sect:abso_O7}}

\begin{figure}
\centering
\includegraphics[width=0.49\textwidth]{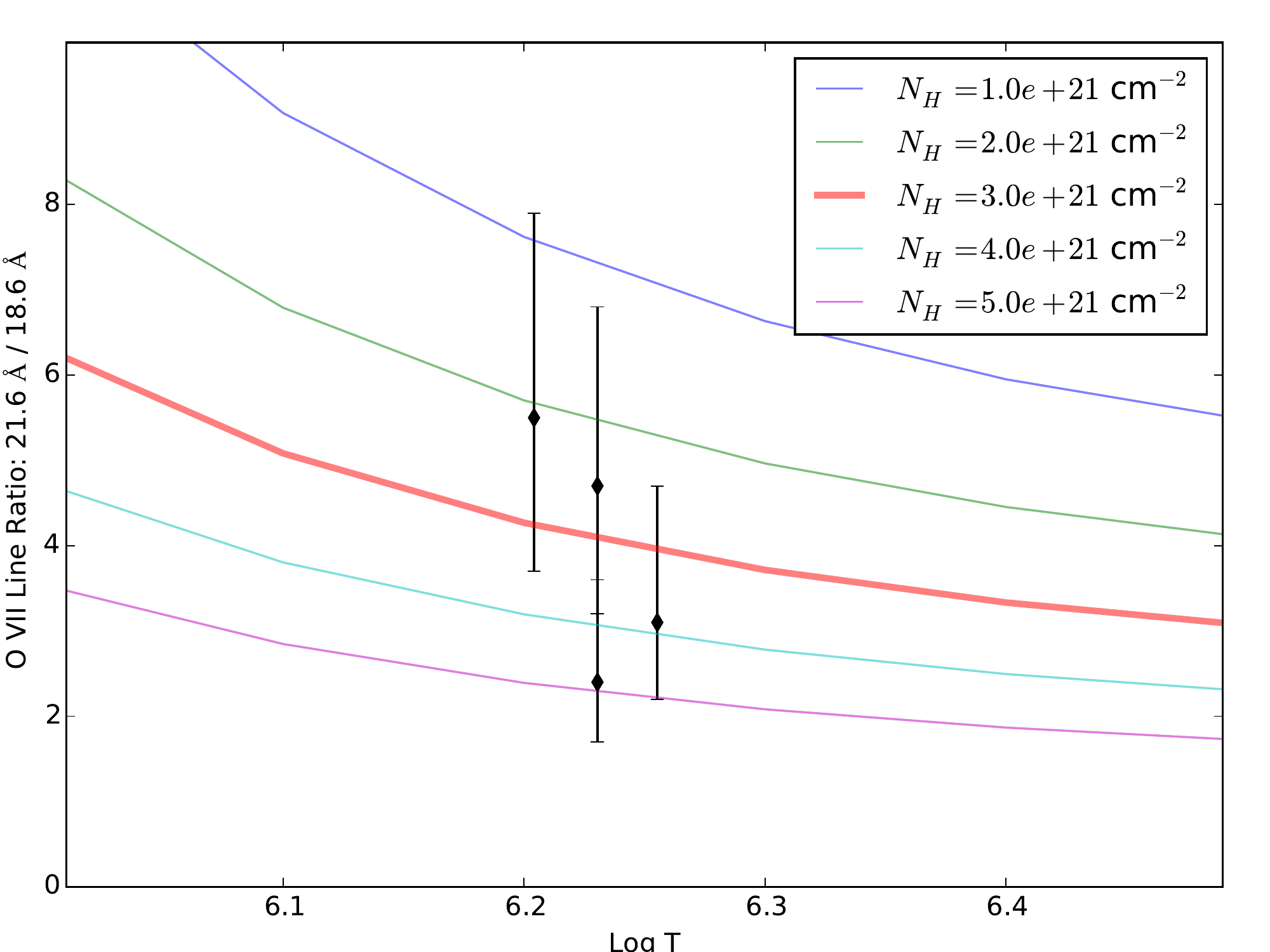}
\caption{Observed ratios for the He-like O~{\sc vii} lines. The four XMM-Newton epochs are plotted
at the temperature of the cool temperature component from the CCD fits. 
Errors indicate 1\,$\sigma$ errors. \label{fig:o7_abs}}
\end{figure}

Dereddening the O~{\sc vii}~r to O~{\sc viii}~Ly\,$\alpha$ ratio
implicitly assumes that the extinction towards 
the O~{\sc vii} and O~{\sc viii} emitting regions is equal. 
However, the deep Chandra grating
spectrum of TW~Hya suggests that this is not necessarily the case
\citep[][]{Brickhouse_2010}. The emitted flux ratio between the
He\,$\alpha$ and He\,$\beta$-like lines of O~{\sc vii} (at $18.6$\,\AA{}
and $21.6$\,\AA, respectively) depends only on the plasma temperature so
that the observed ratio informs us about the absorption towards the 
O~{\sc vii} emitting region provided we 
know the plasma temperature. 
As the temperature estimate, we use the temperature of the coolest
component derived from the CCD spectra (we give the motivation for this choice 
in Sect.~\ref{sect:G_ratio}). Table~\ref{tab:grating_ratios} lists these line 
ratios and Fig.~\ref{fig:o7_abs} compares them with the ratios 
expected for various absorbing column densities and temperatures.
Since these line ratios are compatible with the absorption obtained from 
the CCD spectra, we decided to use $N_H=3\times10^{21}$\,cm$^{-2}$ 
when measuring the soft excess from the oxygen lines.
Figure~\ref{fig:o7_abs} shows the absorption diagnostic for the oxygen {\sc vii} lines.

\end{document}